\documentclass[aps,prd,onecolumn,showpacs,superscriptaddress,showkeys]{revtex4-1}

\usepackage{amssymb}
\usepackage{amsmath}
\usepackage{amsfonts}
\usepackage{graphicx}
\usepackage{color}
\usepackage{xspace}
\usepackage{ulem}
\usepackage{mathtools}
\usepackage{hhline}

\newcommand{\AddrVandy}{Department of Physics and Astronomy, Vanderbilt University, Nashville, TN 37235, USA}

\begin{document}

\title{Stimulated Radiation from Axion Cluster Evolution in Static Spacetimes}

\author{Liang Chen} \email{bqipd@pm.me}\affiliation{\AddrVandy}
\author{Thomas W. Kephart}    \email{tom.kephart@gmail.com}\affiliation{\AddrVandy}

\date{\today}

\begin{abstract}
If their number density is high enough, clusters of axions can decay to photons via stimulated emission. We study both the special and general relativistic corrections to lasing in such dense axion clusters as they evolve in static spacetime of a host object. Our main results are rate equations for the evolution of axion and photon number densities that include these corrections.  We use Schwarzschild spacetime as a detailed example.
\end{abstract}


\maketitle


\section{Introduction}

If axions exist (for reviews of axion physics see \cite{Kim:1986ax,Cheng:1987gp,Raffelt:1990yz}), they are a dark matter (DM) candidate produced in a Bose-Einstein condensate (BEC)
in the early universe by vacuum misalignment \cite{Abbott:1982af,Preskill:1982cy,Dine:1982ah} and/or
by cosmic strings
after inflation \cite{Sikivie:2010bq} where the produced axions relax  into a BEC via
$a+a\rightarrow a+a$
scattering and by the gravitationally scattering of axions. When the temperature
is near the QCD scale ($T\sim 1$GeV),
regions where the axion field $a$ is far from the minimum of its
potential $V(a(T))$
have axion over densities that can decouple from the Hubble flow to
form axion mini clusters. The typical
mass of these objects is roughly $\sim 10^{-14}  M_{\odot}$ \cite{Enander,Fairbairn:2017sil}
and a fraction $f_{mc}\sim 10\%$ of all axions are expected to be in
mini clusters \cite{Fairbairn:2017dmf}. The evolution of axion mini clustered have
been studied in   \cite{Kolb:1993zz,Kolb:1993hw} and more recently in \cite{Braaten:2019knj}.

 Since the central density, radius and total masses of axion miniclusters are known along
their evolutionary stability curves, we can check to
 see if these parameters are in the range where lasing can commence.
If it does   lase, then the cluster
 experiences an immediate (approximate $\delta$-function) change in
those parameters due to the fast mass loss after
 which the cluster  must rearrange itself to regain stability or
quasi-stability.

Gravitationally bound and self bound axion clusters have been studied
by several groups. Relativistic axions
are best expressed in terms of a real scalar field and low energy
axions $E<<m_a$ by an effective complex
field, see   \cite{Guth:2014hsa,Braaten:2019knj,Braaten:2018lmj}.
For a detailed review of for various bound state including
quasi-bound states and including breather like states, see  \cite{Braaten:2019knj}.
 Particles in free space can interact gravitationally or through other mutual interactions to form bound or quasi-bound objects. If the particles are identical bosons in the early universe, then the objects are Bose stars. If the particles are cold or can radiate energy, then they can form a Bose-Einstein condensate (BEC) with properties that differ from low occupation number configurations.   For a recent review of axion cosmology see \cite{Marsh:2015xka}.   Depending on invisible axion model parameters, they can form gravitationally bound clumps called axion stars or self interaction systems called axitons. For an extensive review of bound axion systems with a review of axion field theory, and definitions see again \cite{Braaten:2019knj}. (We assume that the axions considered here to be is states of high occupation number, but not in BECs. They could in principle also be in a large set of overlapping BECs, but not in a single BEC, and in this case they would still behave sufficiently classically that our methods will apply to a good approximation.)

 Let us call all the possible types of bound or quasibound axion configurations axion clusters. Here we will be interested in the evolution of these axion clusters, however we will not need to distinguish between stable or quasi-stable clusters, since we will be studying  axion cluster decay to photons via lasing who's time scale is much shorter than other time scales associated with cluster evolution phenomenon, like relaxation, radiative cooling, etc. Hence the clusters we consider need not even be bound, but could just be dense transients. Hence our focus will be on snap shots of otherwise slowly evolving clusters where we ask if the parameters are right for lasing to commence when we take both general and special relativistic effects into account, i.e., we allow for axion velocities near the speed of light $c$ in curved backgrounds.
 
 The axion field $a$ couples to standard model particles, plus it self interacts via an effective potential $V(a)$, but the coupling of most interest here is to photons via the global chiral anomaly term in the Lagrangian
 $$\frac{\alpha \, c_{\gamma 0}}{8\pi f_a} a F\wedge F$$
 where $\alpha$ is the fine structure constant, $f_a$ is the axion decay constant, $F$ is the electromagnetic field strength and $c_{\gamma 0}$ is an $O(1)$ model dependent constant. From this term we can calculate the decay rate for $a\rightarrow\gamma + \gamma$.
 
 Experimental and cosmological limits have now constrained the mass of the QCD axion to lie in a narrow range $10^{-3} eV > m_a > 10^{-5} eV$. Because axions are copiously produced at rest during the QCD phase transition, they would over close the universe if their mass was lighter than the lower bound and they would have been experimentally detected if they were heavier than the upper bound. Since they are born non-relativistic they are a cold dark matter (CDM) candidate. Studies of density perturbations have shown that CDM can form structures on all scales. Consequently, assuming axions
are the CDM, we expect to find them seeding galaxies or clusters of galaxies, but also perhaps being the seeds of early stars. There are many potential forms of these axion clumps. If they are diffuse the axions will remain non-relativistic. If they become  dense then they can become relativistic. The clumps may or may not be spherically symmetric depending on the environment in which they were formed. If they are sufficiently dense the a curved space metric will be necessary to describe the cluster. If their number density is high enough, then they can lase and hence the resulting photons can be detectable at large distances.

Spherically symmetric non-relativistic lasing axion clusters were first studied in the 1980s   \cite{Kephart:1986vc,Tkachev:1987cd,Kephart:1994uy}. More recently non-spherically symmetric non-relativistic cluster with arbitrary momentum and spacial distributions have also been studied \cite{Chen:2020ufn,Chen:2020yvx}. However, these studies focuses on self-bounded axion cluster. Our purpose here is to study axion cloud bounded by a host object in static space-times. An example is axions moving in a static geometry like Schwarzschild space-time. 

QCD axion clumps with conventional couplings can undergo resonant decay for sufficiently large angular momentum  \cite{Hertzberg:2018zte}.   Radio emission from QCD axions occurs in the context of bose stars if axion-photon coupling is large \cite{Levkov:2020txo}. These are among recent studies on this topic in flat spacetime. There are many standard general relativistic treatment of axion cluster in the literature. \cite{Braaten:2019knj} gives a comprehensive review of axion clusters and includes applications that considered gravitational effect on axion cluster. \cite{Ikeda:2018nhb} shows that laserlike emission from axion clouds exists by numerically solving the problem at classical level in fixed Kerr background. Detailed studies of  axionic instability induced by electromagnetic field deformations of the Kerr-Newman geometry, and periodic bursts of light from superradiant growth are carried out in  \cite{Boskovic:2018lkj}. These studies did the analysis from the perspective of field theory. What sets  \cite{Kephart:1994uy} apart from these studies is that it tackled the problem from statistical model using Boltzmann equation. Therefore, results from \cite{Kephart:1994uy} are potentially applicable to other types of pesudoscalar particle decaying into two photons. Although \cite{Kephart:1994uy} is a statistical model built in flat spacetime, other studies such as \cite{Ikeda:2018nhb} and \cite{Boskovic:2018lkj} suggests curved spacetime may provide more information on this issue. Therefore, we decided to rebuild this model in curved static spacetime. Hence, the methodology implemented here is tailored particularly towards \cite{Kephart:1994uy}. For other approachs of superradiance-blast simulation, please see the studies mentioned above.\\

\section{$a\leftrightarrow\gamma +\gamma$ process in Minkowski spacetime}
This is a brief summary of the formulation of axion cluster lasing using Boltzmann equation in \cite{Kephart:1994uy}.
Two photons emitted by decay of a spin zero particle have the same helicity, as required by angular momentum conservation. The change in the number density of photons of a given helicity $\lambda=\pm1$ within the axion cluster, due to the process $a\leftrightarrow\gamma +\gamma$ in Minkowski spacetime, is
\begin{flalign} \label{nLIPS}
{dn_\lambda \over dt}=&\int dX^{(3)}_\text{LIMS} [f_a(1+f_{1\lambda})(1+f_{2\lambda})-f_{1\lambda}f_{2\lambda}(1+f_a)] \\\nonumber
&\times|M(a\rightarrow\gamma(\lambda)\gamma(\lambda))|^2 ~,
\end{flalign}
where $f_a$, $f_{1\lambda}$ and $f_{2\lambda}$ are the occupation numbers of the axion and the two photons and   $M=M(a\rightarrow\gamma(+)\gamma(+))=M(a\rightarrow\gamma(-)\gamma(-))$ is the decay amplitude determined by the Abelian chiral anomaly\cite{Adler:1969gk,Bell:1969ts} and is related to the spontaneous axion decay constant by
\begin{flalign} \label{axion_decay_constant}
\tau_a^{-1}=\Gamma_a ={1\over{8\pi}} ({1\over{2m_a}}) {1\over2}\sum_{\lambda=\pm}|M(a\rightarrow\gamma(\lambda)\gamma(\lambda))|^2 ~.
\end{flalign}
The three body Lorentz invariant momentum measure is
\begin{flalign} \nonumber
\int dX^{(3)}_\text{LIMS} = & \int {d^3p\over (2\pi)^3 2p^0} \int {d^3k_1\over (2\pi)^3 2k_1^0} \int {d^3k_2\over (2\pi)^3 2k_2^0} \\\nonumber
&\times (2\pi)^4 \delta^{(4)}(p-k_1-k_2) ~.
\end{flalign}
See the Appendix for details of deriving $dX_\text{LIMS}$.
From here one obtains eq. (10) of \cite{Kephart:1994uy}  
\begin{flalign} \label{KW-10}
2k \frac{d f_\lambda ( \vec{k} )  }{dt} = \frac{4 m_{a} \Gamma_{a} }{ \pi } 
\int \frac{d^3  k_1   }{ 2k_1^0 } \frac{d^3  p   }{ 2p^0 }  \delta^4 (p-k-k_1) \\ \nonumber
\times \{  f_a  ( \vec{p} ) [ 1+ f_\lambda ( \vec{k} ) + f_\lambda ( \vec{k_1} ) ] - f_\lambda ( \vec{k} )f_\lambda ( \vec{k_1} ) \} ~.
\end{flalign}\\
which became the starting point for studying the lasing of axion cluster in flat spacetime.\\

\section{$a\leftrightarrow\gamma+\gamma$ process in Static spacetime}
We generalize equation \eqref{nLIPS} to find the change in the number density of photons of a given helicity $=\pm1$ within the axion cluster, due to the process $a\leftrightarrow\gamma + \gamma$ in static spacetime
\begin{flalign}  \nonumber
ds^2= g_{00} dt^2 + g_{ij} dx^i dx^j  ~,
\end{flalign}
 which is
\begin{flalign} \nonumber
{dn_\lambda \over d\tau}=&\int dX^{(3)}_\text{SCMS} [f_a(1+f_{1\lambda})(1+f_{2\lambda})-f_{1\lambda}f_{2\lambda}(1+f_a)] \\\label{agg-reaction}
&\times|M(a\rightarrow\gamma(\lambda)\gamma(\lambda))|^2 ~,
\end{flalign}
where $dX_\text{SCMS}$ is the static spacetime covariant measure, the generalization of  $dX_\text{LIMS}$ in flat spacetime. We give a pedagogical discussion of how we derive and interpret $dX_\text{SCMS}$ in the Appendix.\\

The one body static covariant measure is
\begin{flalign} \nonumber
\int dX_\text{SCMS} = & \int { \sqrt{   ||g_{ij}||} dp^1 dp^2 dp^3 \over  (2\pi)^3  2   \sqrt { - g_{00}    } p^0  } ~,
\end{flalign}
and the three body static covariant measure is
\begin{flalign}\nonumber
     & \int dX^{(3)}_\text{SCMS} \\ \nonumber
 = & \int { \sqrt{   ||g_{ij}||} dp^1 dp^2 dp^3 \over  (2\pi)^3  2   \sqrt { - g_{00}    } p^0  } 
    { \sqrt{   ||g_{ij}||} dk_1^1 dk_1^2 dk_1^3 \over  (2\pi)^3  2   \sqrt { - g_{00}    } k_1^0  }  \\\nonumber
&\times { \sqrt{   ||g_{ij}||} dk_2^1 dk_2^2 dk_2^3 \over  (2\pi)^3  2   \sqrt { - g_{00}    } k_2^0  }
  (2\pi)^4 \delta^{(4)}(p-k_1-k_2) ~.
\end{flalign}

Equation \eqref{axion_decay_constant} describes the axion decay constant $\tau_a$ in a local inertial frame that is comoving with the axion. The relation between the time in the comoving frame $\tau$ with the axion and the coordinate time $t$ in static spacetime is
\begin{flalign}  \nonumber
 {dt\over d\tau}    =& {1\over \sqrt{- g_{00}} } \sqrt{ g_{ij} { p^i p^j  \over m^2 } +  1 } ~.
\end{flalign}
Specifically, in Schwarzschild spacetime
\begin{flalign} \label{schw-metric}
ds^2=-(1-{2M\over r})dt^2 + (1-{2M\over r})^{-1}dr^2 + r^2d\Omega^2  ~, 
\end{flalign}
this becomes
\begin{flalign}  \nonumber
({dt\over d\tau}) =& (1-{2M\over r})^{-1/2} \sqrt{  g_{ij} { p^i p^j  \over m^2 } + 1 }
\end{flalign}
where 
\begin{flalign}  \nonumber
g_{ij}  p^i p^j=& (1-{2M\over r})^{-1} (p^r)^2 + r^2(p^\theta)^2 + r^2\sin^2\theta (p^\phi)^2  
\end{flalign}
is the square of the magnitude of the 3-momentum. 
The decay constant $\tau_a$ in the comoving frame would change to the decay constant  $\tau_{al}$  in lab frame
\begin{flalign}  \nonumber
\tau_{al}= \tau_a {1\over \sqrt{- g_{00}} } \sqrt{ g_{ij} { p^i p^j  \over m^2 } +  1 }   ~,
\end{flalign}
and the decay rate in the lab frame becomes
\begin{flalign}  \nonumber
\Gamma_{al} = \Gamma_a \sqrt{- g_{00}} (  g_{ij} { p^i p^j  \over m^2 } +  1  )^{-1/2}  ~.
\end{flalign}

In a static spacetime, the number density  given by
\begin{flalign} \nonumber
n(x^\alpha)=\int f(p^i,x^\alpha) { d^3p \over (2\pi)^3 } = \int f(p^i,x^\alpha) { \sqrt{ ||g_{ij}|| } \over (2\pi)^3 } dp^1 dp^2 dp^3  ~,
\end{flalign}
where $|g_{ij}|$ is the determinant of the 3-surface metric
\begin{flalign} \nonumber ds^2= g_{ij}dx^i dx^j  \end{flalign}
at constant coordinate time $t$. The total number of the particle is
\begin{flalign} \nonumber
N(t)=\int n(x^\alpha)   d^3x = \int n(x^\alpha) \sqrt{ ||g_{ij}|| }dx^1 dx^2 dx^3 ~,
\end{flalign}
In static spacetime, the rate of change in number density is related to the rate of change in occupation number by
\begin{flalign} \nonumber
{dn_\lambda \over dt}=&\int 2  \sqrt { - g_{00} }  k^0 {df_\lambda \over dt} {\sqrt{ ||g_{ij}|| } dk^1 dk^2 dk^3 \over (2\pi)^3 2 \sqrt { - g_{00} } k^0 } ~.
\end{flalign}
In Minkowski spacetime $k^0=k$, but the magnitude of energy and momentum are no longer equal in general  in a static spacetime. Instead, we have a static energy-momentum relation for photon:
\begin{flalign}  \label{photon-e-p}
g_{ij}  k^i k^j =&  - g_{00} ({k^0})^2 ~,
\end{flalign}
and for massive particle, the corresponding relation is
\begin{flalign}  \label{mass-e-p}
m^2 + g_{ij}  p^i p^j  =&  - g_{00} (p^0)^2  ~.
\end{flalign}\\

Multiplying  equation \eqref{agg-reaction} by ${d\tau\over dt}$, we have an equation for the rate of change in the number density of photons measured by coordinate time $t$,
\begin{flalign} \nonumber
{dn_\lambda \over dt}=&\int dX^{(3)}_\text{SCMS} \{ [f_a[1+f_\lambda(k^i) +f_\lambda(k_1^i)] \\\nonumber
& - f_\lambda(k^i)  f_\lambda(k_1^i) \}  \times 16\pi m_a \Gamma_a {d\tau\over dt} ~.
\end{flalign}
Writing  $dn_\lambda \over dt$ in terms of $f_a$ we have one body measure $dX_\text{SCMS}$  on the LHS of this equation that we use to cancel one of the three measures in $dX^{(3)}_\text{SCMS}$ on the RHS. There remains two measures, one for the axion and one for a photon, plus the $\delta$ function,
\begin{flalign} \nonumber
2  \sqrt { - g_{00} }  k^0 {df_\lambda \over dt} =&  \int { \sqrt{   ||g_{ij}||} dp^1 dp^2 dp^3 \over  (2\pi)^3  2   \sqrt { - g_{00}    } p^0  } 
    { \sqrt{   ||g_{ij}||} dk_1^1 dk_1^2 dk_1^3 \over  (2\pi)^3  2   \sqrt { - g_{00}    } k_1^0  }   \\\nonumber
& \times (2\pi)^4 \delta^4 (p^\alpha-k^\alpha-k_1^\alpha)  \\\nonumber
& \times \{ [f_a[1+f_\lambda(k^i) +f_\lambda(k_1^i)] - f_\lambda(k^i)  f_\lambda(k_1^i) \} \\\nonumber
& \times 16\pi m_a \Gamma_a \sqrt{- g_{00}} (  g_{ij} { p^i p^j  \over m_a^2 } +  1  )^{-1/2}   ~.
\end{flalign}
After simplification, we obtained the  evolution equation  
\begin{flalign} \nonumber
2     k^0 \frac{d f_\lambda }{dt} =& \frac{4 m_{a} \Gamma_{a} }{ \pi } 
\int { \sqrt{   ||g_{ij}||} dp^1 dp^2 dp^3 \over   2   \sqrt { - g_{00}    } p^0  } 
    { \sqrt{   ||g_{ij}||} dk_1^1 dk_1^2 dk_1^3 \over   2   \sqrt { - g_{00}    } k_1^0  }   \\\nonumber
& \times \{ [f_a[1+f_\lambda(k^i) +f_\lambda(k_1^i)] - f_\lambda(k^i)  f_\lambda(k_1^i) \} \\\label{evolution}
& \times \delta^4 (p^\alpha-k^\alpha-k_1^\alpha) (  g_{ij} { p^i p^j  \over m_a^2 } +  1  )^{-1/2} ~.
\end{flalign}
which resembles equation  \eqref{KW-10}, and reduces to it in the flat spacetime limit.

\section{Momentum Space Integrations }
The momentum space integrations needed for the evaluation of \eqref{evolution} are rather tedious and can be found in the Appendix.  Specifically, doing the $k_1$ integration first leads to 
\begin{flalign} \nonumber
&  2k^0 \frac{d f_\lambda ( k^i )  }{dt} \\ \nonumber
=& \frac{4 m_a \Gamma_{a} }{ \pi  ( - g_{00} )  }    \int  {m_a\over p^0 } { \sqrt{   ||g_{ij}||} dp^1 dp^2 dp^3 \over   2  \sqrt{  - g_{00} } p^0   }  \\ \nonumber
&\times  {   1 \over   2    \sqrt{ g_{ij}  (p^i-k^i) (p^j-k^j)  }  }  \\ \nonumber
&\times \delta [  p^0 - k^0- \sqrt{ g_{ij}  (p^i-k^i) (p^j-k^j) \over  - g_{00} }  ]  \\ \label{rate-equa-1}
&\times \{   f_a  ( p^i ) [ 1+ f_\lambda ( k^i ) + f_\lambda ( p^i-k^i ) ] - f_\lambda ( k^i )f_\lambda ( p^i-k^i ) \} ~.
\end{flalign}
and then we assume that the occupation numbers of both axion and photon are isotropic,
\begin{flalign} \nonumber
f_a  ( p^i ) =  f_a  ( \sqrt{ g_{ij} p^i p^j } )  ~,  \qquad
f_\lambda ( k^i ) =  f_\lambda ( \sqrt{ g_{ij} k^i k^j } ) ~,
\end{flalign}
and the $p$ integration given in the Appendix results in
\begin{flalign} \label{rate-equa-5}
&    \frac{d f_\lambda ( \sqrt{ g_{ij} k^i k^j }   )  }{dt}   \\ \nonumber
= &  { m_a \Gamma_{a} \sqrt{   - g_{00}  }  \over    g_{ij} k^i k^j   }   \int  {m_a\over p^0    } 
d p^0   \times  \{   f_a  ( \sqrt{ g_{ij} p^i p^j }    )   \\ \nonumber
&   [ 1+ f_\lambda ( \sqrt{ g_{ij} k^i k^j }     )   + f_\lambda \left(\sqrt{ - g_{00} }  ( p^0   - k^0 )   \right)  ]  \\  \nonumber
 &  -  f_\lambda (  \sqrt{ g_{ij} k^i k^j }      )   f_\lambda \left( \sqrt{ - g_{00} }  ( p^0   - k^0 )   \right)  \} ~.
\end{flalign}
where we have a factor of $\sqrt{   - g_{00}  }$ from gravitational redshift which corrects the time difference between the clock in the lab frame and the clock at the location the axion. The factor $ {m_a\over p^0}$ is due to the special relativistic  correction.\\

\section{Evolution equations}
In the Appendix, we find that at event ${\mathcal P_0}$, there are upper and lower limits on the $0^{th}$ component of photon momentum. This is true for every event  at any point in spacetime. The bounds are
\begin{flalign}  \nonumber
 p^0     \geq &     k^0  + { m_a^2 \over 4  k^0  (-g_{00}) }  \\ \nonumber
 k^0_\text{max/min}     =& { 1 \over  2 } [\  p^0 \pm \sqrt{ (  p^0)^2 -  { m_a^2 \over -g_{00} }  } \ ] 
    \\ \nonumber
=&   { 1   \over  2 }  ( p^0 \pm  \sqrt{g_{ij} p^i p^j \over -g_{00} } )     \\ \nonumber
\sqrt{ g_{ij} k^i k^j }_\text{max/min}= &   { 1\over2 }  ( \sqrt{-g_{00}} p^0 \pm  \sqrt{g_{ij} p^i p^j  } )~.
\end{flalign}
Switching from variable $\sqrt{ g_{ij} p^i p^j }$ to $p^0$, the evolution equation becomes
\begin{flalign} \nonumber
&    \frac{d f_\lambda ( \sqrt{ g_{ij} k^i k^j }   )  }{dt}   \\ \nonumber
= &  { m_a \Gamma_{a} \sqrt{   - g_{00}  }  \over    g_{ij} k^i k^j   }   \int_{ k^0  - {m_a^2 \over 4  k^0  g_{00}}   }    {m_a\over p^0    }  d p^0 \{   f_a  \left( \sqrt{  ( - g_{00}) (p^0)^2 - m_a^2 }     \right)   \\ \nonumber
& \times   [ 1+ f_\lambda ( \sqrt{ g_{ij} k^i k^j }     )   + f_\lambda \left(\sqrt{ - g_{00} } ( p^0 - k^0 ) \right)  ]   \\ \nonumber
&       -  f_\lambda (  \sqrt{ g_{ij} k^i k^j }      )   f_\lambda \left( \sqrt{ - g_{00} } ( p^0 - k^0 )   \right)  \} ~.
\end{flalign}
 Note that $p^0=k^0 + k_1^0$ and $\sqrt{   - g_{00}  } k_1^0 = \sqrt{ g_{ij} k_1^i k_1^j }  $. The $p^0$ integral can be rewritten as an  integral over $\sqrt{ g_{ij} k_1^i k_1^j }$,
\begin{flalign} \nonumber
&    \frac{d f_\lambda ( \sqrt{ g_{ij} k^i k^j }   )  }{dt}   \\ \nonumber
= &  { m_a \Gamma_{a} \sqrt{   - g_{00}  }  \over    g_{ij} k^i k^j   }   \int_{  m_a^2 \over 4 \sqrt{ g_{ij} k^i k^j }   }  
  {m_a   d ( \sqrt{ g_{ij} k_1^i k_1^j })   \over  \sqrt{ g_{ij} k^i k^j } +  \sqrt{ g_{ij} k_1^i k_1^j }   }     \\ \nonumber
& \times  \{   f_a  \left( \sqrt{   ( \sqrt{ g_{ij} k^i k^j } + \sqrt{ g_{ij} k_1^i k_1^j })^2 - m_a^2 }    \right)     \\ \nonumber
& \times [ 1+ f_\lambda ( \sqrt{ g_{ij} k^i k^j }     )   + f_\lambda ( \sqrt{ g_{ij} k_1^i k_1^j }   )  ]  \\ \nonumber
&   -  f_\lambda (  \sqrt{ g_{ij} k^i k^j }      )   f_\lambda ( \sqrt{ g_{ij} k_1^i k_1^j }   )  \} ~.
\end{flalign}
We convert the LHS of this equation into the changing rate of the number density by integrating over the momentum space,
\begin{flalign} \nonumber
& \frac{d n_\lambda }{dt}    =  \int   \frac{d f_\lambda ( \sqrt{ g_{ij} k^i k^j }  )  }{dt}  { g_{ij} k^i k^j \over 2\pi^2 } 
    d( \sqrt{ g_{ij} k^i k^j } )     ~.
\end{flalign}
The rate of change in photon number density is then
\begin{flalign} \nonumber
&    \frac{d n_\lambda  }{dt}  
=   { m_a \Gamma_{a} \sqrt{   - g_{00}  }  \over    2\pi^2  }  \int  \int_{ k^0  - {m_a^2 \over 4  k^0  g_{00}}   }    {m_a\over p^0    }   \{   f_a  \left( \sqrt{  ( - g_{00}) (p^0)^2 - m_a^2 }    \right)   \\ \nonumber
& \times   [ 1+ f_\lambda ( \sqrt{ g_{ij} k^i k^j }     )   + f_\lambda ( \sqrt{ - g_{00} } p^0 - \sqrt{ g_{ij} k^i k^j }   )  ]   \\ \nonumber
&       -  f_\lambda (  \sqrt{ g_{ij} k^i k^j }      )   f_\lambda ( \sqrt{ - g_{00} } p^0 - \sqrt{ g_{ij} k^i k^j }   )  \}  \\ \nonumber
& \times  d p^0 d( \sqrt{ g_{ij} k^i k^j } )  ~,
\end{flalign}
or alternatively
\begin{flalign} \label{rate-equa-2}
&   \frac{d n_\lambda  }{dt}   
=   { m_a \Gamma_{a} \sqrt{   - g_{00}  }  \over   2\pi^2   }  \int  \int_{  m_a^2 \over 4 \sqrt{ g_{ij} k^i k^j }   }   
       {m_a  d ( \sqrt{ g_{ij} k_1^i k_1^j })   d( \sqrt{ g_{ij} k^i k^j } )    \over  \sqrt{ g_{ij} k^i k^j } +  \sqrt{ g_{ij} k_1^i k_1^j }   }  \\ \nonumber
& \times  \{   f_a  \left( \sqrt{   ( \sqrt{ g_{ij} k^i k^j } + \sqrt{ g_{ij} k_1^i k_1^j })^2 - m_a^2 }   \right)  [ 1+ f_\lambda ( \sqrt{ g_{ij} k^i k^j }     )    \\ \nonumber
&    + f_\lambda ( \sqrt{ g_{ij} k_1^i k_1^j }   )  ]     -  f_\lambda (  \sqrt{ g_{ij} k^i k^j }      )   f_\lambda ( \sqrt{ g_{ij} k_1^i k_1^j }   )  \}    ~.
\end{flalign}

We assume that the various dependences of the axion occupation number are separable, and of the form
\begin{flalign} \nonumber
& f_a(\sqrt{ g_{ij} p^i p^j }, r,t ) \\ \nonumber
=& \Theta( p_\text{max} - \sqrt{ g_{ij} p^i p^j } ) [ f_{ac}(t)\Theta(r_+ - r)\Theta(r-r_-) \\ \label{axion-occu}
& + f_{ad}(t) d(r) ] ~.
\end{flalign}
We let $r_+ \sim r_-$ both be far beyond the event horizon if the host object is a black hole, with the small distortion $d(r)$ in the region $(r_-\leq r\leq r_+)$  away  from a uniform  distribution. In the Appendix, we find that the maximum momentum of an axion is
$$   p_\text{max}  = m_a \beta' =    m_a \sqrt{ {- g_{00}( r_+)\over - g_{00} (r_-)} - 1 }  ~.$$

Since we set the maximum axion momentum to be $m_a\beta'$, according to the bounds mentioned in the previous section, extremes of photon momentum become
\begin{flalign}  \nonumber
\sqrt{ g_{ij} k^i k^j }_\pm =  &  { 1\over2 }  ( \sqrt{-g_{00}} p^0 \pm  \sqrt{g_{ij} p^i p^j  } )  \bigg|_{ \sqrt{g_{ij} p^i p^j  } = m_a\beta'}  \\  \nonumber
=  &  { 1\over2 }   [ \sqrt{ m_a^2 + (m_a\beta')^2 }   \pm  \sqrt{ (m_a\beta')^2   } ]     \\  \nonumber
 = &   { m_a \over  2 }  (   \sqrt{1+\beta'^2} \pm  \beta'  )         ~.
\end{flalign}

This is somewhat different from the extreme photon momentum values in \cite{Kephart:1994uy}, because there the escape velocity was defined in the non-relativistic sense.\\
Suppose that the photon occupation number which correspondences to that of axions \eqref{axion-occu}, is
\begin{flalign}  \nonumber
& f_\lambda(\sqrt{ g_{ij} k^i k^j }, r, t )   \\  \label{photon-occu} 
 = & [ f_{\lambda c}(t)\Theta(r_+ - r)\Theta(r-r_-) + f_{\lambda d}(t) d(r) ] \\  \nonumber
 &\times \Theta( \sqrt{ g_{ij} k^i k^j }_+ -  \sqrt{ g_{ij} k^i k^j } ) \Theta( \sqrt{ g_{ij} k^i k^j } - \sqrt{ g_{ij} k^i k^j }_- ) ~.
\end{flalign}
The volume of the shell $ \Theta( \sqrt{ g_{ij} k^i k^j }_+ -  \sqrt{ g_{ij} k^i k^j } ) \Theta( \sqrt{ g_{ij} k^i k^j } - \sqrt{ g_{ij} k^i k^j }_- ) $ is
\begin{flalign}  \nonumber
 V_k  \approx &  4\pi [ {1\over2}  ( \sqrt{ g_{ij} k^i k^j }_+ + \sqrt{ g_{ij} k^i k^j }_- ) ]^2  \\  \nonumber
 & \times  ( \sqrt{ g_{ij} k^i k^j }_+ - \sqrt{ g_{ij} k^i k^j }_- )   = \pi m_a^3 \beta'(1+\beta'^2)  ~.
\end{flalign}

The $k$ and $k_1$ integrations of \eqref{rate-equa-2} are both long and tedious and so have been relegated to the Appendix. They result in the following expression
\begin{flalign} \nonumber
&  \frac{d n_\lambda   }{dt} =  { m_a \Gamma_{a} \over   2\pi^2   }  \sqrt{   - g_{00}  }    \\ \nonumber
&\times   \{  {  m_a^2 \beta'^3 \over 3}  [  f_{ac}(t)\Theta(r_+ - r)\Theta(r-r_-)   + f_{ad}(t) d(r)    \\ \nonumber
 &     +  2 f_{ac}(t)  f_{\lambda c}(t)\Theta(r_+ - r)\Theta(r-r_-)    \\ \nonumber
 & + 2  (  f_{ac}(t)  f_{\lambda d}(t)  +  f_{ad}(t)  f_{\lambda c}(t)  ) d(r)   \\ \nonumber
& -   f_{\lambda c}^2(t)\Theta(r_+ - r)\Theta(r-r_-) - 2 f_{\lambda c}(t) f_{\lambda d}(t) d(r)   ]  \\ \label{DnPhotonDt}
&  -    { m_a^2\beta'^2 \over  2 }       [ f_{\lambda c}^2(t)\Theta(r_+ - r)\Theta(r-r_-) + 2 f_{\lambda c}(t) f_{\lambda d}(t) d(r)   ]   \}   ~,
\end{flalign}
where the subscripts $c$ and $d$ refer to the flat and distorted part of the spectrum. (See the next two sections for more discussion of the distortion function $d(r)$.)
This result will be used to analyze the Schwarzschild case in the next section. Later we  apply the results to a specific model and extract
the distortion caused by the  relativistic corrections of the flat space case. Equation \eqref{DnPhotonDt} is one of the main results of this work.

 The rate of change in axion number density is the opposite of that of photon, and is
\begin{flalign} \nonumber
&  \frac{d n_a   }{dt} = -{1\over2}\sum_{\lambda=\pm} \frac{d n_\lambda    }{dt}   ~.
\end{flalign}
 as we find from \eqref{DnPhotonDt}.

\section{Setup of simple cluster model}
The rate equations we have just developed applies to all static spacetime metrics, includes those from metric based theories other than general relativity. E.g., there could be static host objects in other metric based gravitational theory that allow spontaneous particle creation which serve as a source of the axion production. But for now we  set the question of axion production aside and focus on the gravitational correction to stimulated radiation in axion clusters.

The major difference between this model and \cite{Kephart:1994uy} is that here the axion self gravity is ignored but the gravity from the host objects(star, black hole, etc) is taken into account. If these axions were produced by perturbative black hole processes, we assume the total energy in axions  be much smaller than the mass of the black hole. 

We want to investigate static stationary spacetime where
\begin{flalign} \nonumber
g_{00}=g_{00}(r) \qquad g_{ij}=g_{ij}(r,\theta)    \qquad
\lim_{r\rightarrow\infty} g_{\mu\nu} = \eta_{\mu\nu} ~.
\end{flalign}
Schwarzschild and Reissner-Nordstr{\"o}m metric belong to this category.
Because of the factor $\sqrt{-g_{00}}$, it is difficult to maintain a uniform distribution for photons along the radial direction. Thus we introduce a small distortion function $d(r)$ in the following calculations.\\

The axion number density at event $x^\alpha$ is an integration of $f_a(\sqrt{ g_{ij} p^i p^j }, r)$ over $p^i$,
\begin{flalign}  \nonumber
n_a(t, x^i )   =&  \int_0^{m_a\beta'}  \Theta( p_\text{max} - \sqrt{ g_{ij} p^i p^j } ) 
  { g_{ij} k^i k^j \over 2\pi^2 }     d( \sqrt{ g_{ij} k^i k^j } )  \\  \nonumber
  & \times   [ f_{ac}(t)\Theta(r_+ - r)\Theta(r-r_-)   + f_{ad}(t) d(r) ]     \\  \nonumber
= & { (m_a\beta')^3 \over 6\pi^2 }    [ f_{ac}(t)\Theta(r_+ - r)\Theta(r-r_-)   + f_{ad}(t) d(r) ]  \\  \nonumber
= &   [ n_{ac}(t)\Theta(r_+ - r)\Theta(r-r_-)   + n_{ad}(t) d(r) ]~.
\end{flalign}
The   occupation numbers can be converted to  number densities with
\begin{flalign}  \nonumber
 f_{ac}(t)  = & { 6\pi^2   \over (m_a\beta')^3    } n_{ac}(t)      \\  \nonumber
 f_{ad}(t)  = & { 6\pi^2   \over (m_a\beta')^3    } n_{ad}(t)      
\end{flalign}
The number density of photon 
\begin{flalign}  \label{distorted.n.photon}
n_\lambda(t, x^i )   = &  n_{\lambda c}(t)\Theta(r_+ - r)\Theta(r-r_-) + n_{\lambda d}(t) d(r)
\end{flalign}
is an integration of photon occupation number $f_\lambda$ over $k^i$. The coefficients of occupation number and number density are related by
\begin{flalign}  \nonumber
 f_{\lambda c}(t) = &{ (2\pi)^3 n_{\lambda c}(t) \over V_k } = { 8\pi^2 \over m_a^3 \beta' } n_{\lambda c}(t)  \\  \nonumber
f_{\lambda d}(t) = &{ (2\pi)^3 n_{\lambda d}(t) \over V_k } = { 8\pi^2 \over m_a^3 \beta' } n_{\lambda d}(t) ~,
\end{flalign}
which we note is different from the axion relations.

\section{Radial distribution approximation}
Because of the factor $ \sqrt{ - g_{00} }$, both axion and photon can not maintain a uniform radial distribution and it is this factor that causes the distortion $d(r)$ to arise in those quantities. We can assume that the distortion is the displacement of $ \sqrt{ - g_{00} }$ from 1.
\begin{flalign} \nonumber
\sqrt{ - g_{00} } = 1 + d(r)  ~.
\end{flalign}
In the case of Schwarzschild spacetime, the Maclaurin series for the correction factor $ \sqrt{ - g_{00} }$ is
\begin{flalign} \nonumber
 \sqrt{ 1-{2M\over r} } =& 1 -  {1\over2} {2M\over r} -  {1\over8}({2M\over r} )^2 -... \end{flalign}
In the case of Reissner-Nordstr{\"o}m spacetime it is
\begin{flalign} \nonumber
 \sqrt{ 1-{2M\over r} + {Q^2\over r^2} } =& 1 -  {1\over2}  {2M\over r} + [ {1\over2} {Q^2\over r^2} -  {1\over8}({2M\over r} )^2 ] -... 
\end{flalign}
From the Maclaurin series, we can see that all the $r$ dependent terms contribute to the unevenness of the distribution, i.e., to $d(r)$. \\

Returning to the general form, we consider the time derivative of \eqref{distorted.n.photon}. Neglecting terms which are second or higher order of $d(r)$, we have
\begin{flalign} \nonumber
&  \frac{d n_\lambda   }{dt} = {d n_{\lambda c}(t) \over dt} \Theta(r_+ - r)\Theta(r-r_-) +  {d n_{\lambda d}(t) \over dt} d(r) \\ \nonumber
=&  {m_a  \Gamma_a \over 2\pi ^2  }[ 1 + d(r) ]  \{  {  m_a^2 \beta'^3 \over 3} 
                [  f_{ac}(t)\Theta(r_+ - r)\Theta(r-r_-)   + f_{ad}(t) d(r)    \\ \nonumber
 &     +  2 f_{ac}(t)  f_{\lambda c}(t)\Theta(r_+ - r)\Theta(r-r_-)    \\ \nonumber
 & + 2  (  f_{ac}(t)  f_{\lambda d}(t)  +  f_{ad}(t)  f_{\lambda c}(t)  ) d(r)   \\ \nonumber
& -   f_{\lambda c}^2(t)\Theta(r_+ - r)\Theta(r-r_-) - 2 f_{\lambda c}(t) f_{\lambda d}(t) d(r)   ]  \\ \nonumber
&  -     { m_a^2\beta'^2 \over  2 }       [ f_{\lambda c}^2(t)\Theta(r_+ - r)\Theta(r-r_-) + 2 f_{\lambda c}(t) f_{\lambda d}(t) d(r)   ]   \}   ~.
\end{flalign}
which defines the uniform and distorted parts of the number density,  $n_{\lambda c}(t)$ and $n_{\lambda d}(t)$ respectively, as well as the 
 uniform and distorted parts of the of the corresponding occupation number.

We now match terms according to whether the radial distribution is uniform $\Theta(r_+ - r)\Theta(r-r_-)$ or distorted $d(r)$. The uniform part of the photon distribution is
\begin{flalign} \nonumber
    {d n_{\lambda c}(t) \over dt}  
=  {m_a  \Gamma_a \over 2\pi ^2  } \{   {  m_a^2 \beta'^3 \over 3} \nonumber
     [  f_{ac}(t)   +  2 f_{ac}(t)  f_{\lambda c}(t)   -   f_{\lambda c}^2(t)  ] -     { 1 \over  2 }    m_a^2\beta'^2   f_{\lambda c}(t)^2   \}  ~.
\end{flalign}
while the deformed part is
\begin{flalign} \nonumber
& {d n_{\lambda d}(t) \over dt}  \\ \nonumber
=& {m_a  \Gamma_a \over 2\pi ^2  } \{ {  m_a^2 \beta'^3 \over 3} \\ \nonumber
& \times  [  f_{ac}(t)      + f_{ad}(t)  + 2 f_{ac}(t)  f_{\lambda c}(t)  + 2   f_{ac}(t)  f_{\lambda d}(t)    \\ \nonumber
&  + 2 f_{ad}(t)  f_{\lambda c}(t) -  f_{\lambda c}^2(t)  - 2 f_{\lambda c}(t) f_{\lambda d}(t)    ]   \\ \nonumber
&    -     { m_a^2\beta'^2 \over  2 }       [  f_{\lambda c}^2(t)  + 2 f_{\lambda c}(t) f_{\lambda d}(t)     ]   \}   ~.
\end{flalign}
Finally we replace all the occupation number coefficents with number density coefficents and simplify to find
\begin{flalign} \nonumber
    {d n_{\lambda c}  \over dt}   \nonumber
=   \Gamma_a  (   n_{ac}     +     { 16 \pi^2   \over m_a^3 \beta' } n_{ac}n_{\lambda c}   -    { 32 \pi^2  \beta'   \over 3 m_a^3  }  n_{\lambda c}^2    -          {  16\pi^2   \over m_a^3  }  n_{\lambda c}^2 ) 
\end{flalign}
which is the   same   as   equation (32) of \cite{Kephart:1994uy}, as expected. In addition, we have the simplified equation that accounts for radial distortion,
\begin{flalign} \nonumber
& {d n_{\lambda d}(t) \over dt}  \\ \nonumber
=&  \Gamma_a  (   n_{ac}   +    n_{ad}  +     { 16\pi^2   \over m_a^3 \beta' } n_{ac}n_{\lambda c} +     { 16\pi^2   \over m_a^3 \beta' }     n_{ac}n_{\lambda d}  +    { 16\pi^2    \over m_a^3 \beta' } n_{\lambda c} n_{ad} \\ \nonumber
&     -  { 32 \pi^2    \beta' \over 3 m_a^3  }n_{\lambda c}^2  
       -   { 64 \pi^2  \beta'    \over  3 m_a^3  } {   n_{\lambda c} n_{\lambda d} }   
  -        { 16\pi^2  \over m_a^3}  n_{\lambda c}^2   -  { 32\pi^2 \over m_a^3   }   n_{\lambda c}    n_{\lambda d}    ) ~.
\end{flalign}\\

\section{Surface loss and total photon density}
 The surface loss of photons at $r=r_+$ is
\begin{flalign}  \nonumber
& (d n_\lambda)_{r_+\text{surface loss}}   \\ \nonumber 
= &   {1\over2} \times \frac{ - d N_\lambda  }{V}  = - {1\over2}   \times \frac{  n_\lambda d V  }{ \int \sqrt{|g_{ij}|} dx^i dx^2 dx^3 }     ~,
\end{flalign}
where the factor $1\over2$ accounts for the probability that in the tangent space of an event at the surface, the momentum of the photon has positive radial component. In general, the surface loss rate $\Gamma_s$ (at both $r_+$ and $r_-$) is proportional to the number density,
\begin{flalign}  \nonumber
& ( \frac{d n_\lambda  }{dt}   )_{r_+\text{surface loss}} + ( \frac{d n_\lambda  }{dt}   )_{r_-\text{surface loss}}  = - \Gamma_s n_\lambda ~.
\end{flalign}

For Schwarzschild spacetime,
\begin{flalign}  \nonumber
& (d n_\lambda)_{r_+\text{surface loss}}   \\ \nonumber 
=& - \frac{  n_\lambda S  c\, (dt)   }{\int  \sqrt{(1-{2M\over r})^{-1} r^2 r^2\sin^2\theta} dr d\theta d\phi } \times \frac{1}{2} \\ \nonumber 
=& - \frac{  n_\lambda S  c\, (dt)   }{ 4\pi  \int_{r_-}^{r_+}  {r^2\over \sqrt{1-{2M\over r}} } dr  } \times \frac{1}{2}    \\ \nonumber 
\approx & - n_\lambda \frac{4\pi r_+^2}{\frac{4\pi r_+^3}{3} - {4\pi r_-^3 \over 3} } \,c\, (dt)   \times \frac{1}{2}   ~.
\end{flalign}
Here since $r\gg 2M$,   (working to lowest order in $d(r)$) we used the approximation $\sqrt{1-{2M\over r}}\sim1$. Then the surface loss at $r_+$ is
\begin{flalign}  \nonumber
& ( \frac{d n_\lambda  }{dt}   )_{r_+\text{surface loss}} =- \frac{3c r_+^2 n_\lambda}{ 2(r_+^3 - r_-^3) }  \\ \nonumber
=&   -  \frac{3c r_+^2 }{ 2(r_+^3 - r_-^3) }       [ n_{\lambda c}(t)\Theta(R-r)  + n_{\lambda d}(t) d(r) ]  ~.
\end{flalign}\\

The surface loss of photon at $r=r_-$ can be neglected because photons would go back to the cluster unless being captured by the host. With any nonzero angular momentum, $r_-\gg 2M$ limits the possibility of incidents that photon falling into the host. Define the surface loss rate,
\begin{flalign}  \nonumber
& \Gamma_s =  -  \frac{3c r_+^2 }{ 2(r_+^3 - r_-^3) }   ~.
\end{flalign}
Surface loss rate for Reissner-Nordstr{\"o}m spacetime can be obtained following similar procedures.\\

With surface loss included, the photon number density rate equation becomes
\begin{flalign} \nonumber
&    {d n_{\lambda c}  \over dt}  \\ \nonumber
=&   \Gamma_a [  n_{ac}     +     { 16 \pi^2   \over m_a^3 \beta' } n_{ac}n_{\lambda c}   -    { 32 \pi^2     \over 3 m_a^3  }  (\beta'+{3\over2}) n_{\lambda c}^2 ]    -\Gamma_s n_{\lambda c}  ~.
\end{flalign}
\begin{flalign} \nonumber
&    {d n_{\lambda d}  \over dt}  \\ \nonumber
=&  \Gamma_a [   n_{ac}   +    n_{ad}  +     { 16\pi^2   \over m_a^3 \beta' } ( n_{ac}n_{\lambda c} +      n_{ac}n_{\lambda d}  +     n_{\lambda c} n_{ad} )\\ \nonumber
&     -  { 32 \pi^2     \over 3 m_a^3  } (\beta' + {3\over2}) n_{\lambda c}^2  
       -   { 64 \pi^2  \beta'    \over  3 m_a^3  } (\beta' + {3\over2}) {   n_{\lambda c} n_{\lambda d} } ]  -\Gamma_s n_{\lambda d} ~.
\end{flalign}\\
Assuming that all  reactions create and/or annihilate equal number of photons from each helicity state, $n_{+c}=n_{-c}$ and $n_{+d}=n_{-d}$. Hence $n_{\lambda c}={1\over2} n_{\gamma c}$ and $n_{\lambda d}={1\over2} n_{\gamma d}$, and we find

\begin{flalign} \nonumber
    {d n_{\gamma c}  \over dt}  
=&   \Gamma_a [ 2 n_{ac}     +     { 16 \pi^2   \over m_a^3 \beta' } n_{ac}n_{\gamma c}   -    { 16 \pi^2     \over 3 m_a^3  }  (\beta'+{3\over2}) n_{\gamma c}^2 ]    -\Gamma_s n_{\gamma c}  ~.
\end{flalign}
and
\begin{flalign} \nonumber  
  {d n_{\gamma d}  \over dt}  =&  \Gamma_a [  2 n_{ac}   +  2  n_{ad}  +     { 16\pi^2   \over m_a^3 \beta' } ( n_{ac}n_{\gamma c} +      n_{ac}n_{\gamma d}  +     n_{\gamma c} n_{ad} )\\ \nonumber
&     -  { 16 \pi^2     \over 3 m_a^3  } (\beta' + {3\over2}) n_{\gamma c}^2  
       -   { 32 \pi^2      \over  3 m_a^3  } (\beta' + {3\over2}) {   n_{\gamma c} n_{\gamma d} } ]  -\Gamma_s n_{\gamma d} ~.
\end{flalign}\\

The uniform and distorted axion number density rate equations are minus one half of the photon number density rate equation, excluding sterile axions we find
\begin{flalign} \nonumber
    {d n_{ac}  \over dt} 
=  - \Gamma_a [ + n_{ac}     +     { 8 \pi^2   \over m_a^3 \beta' } n_{ac}n_{\gamma c}  -    { 8 \pi^2     \over 3 m_a^3  }  \beta' n_{\gamma c}^2 ]    ~.
\end{flalign}
and
\begin{flalign} \nonumber
   {d n_{ad}  \over dt}   
=&  -\Gamma_a [  + n_{ac}   +    n_{ad}  +    { 8\pi^2   \over m_a^3 \beta' } ( n_{ac}n_{\gamma c} +      n_{ac}n_{\gamma d}  +     n_{\gamma c} n_{ad} )\\ \nonumber
&     -  { 8 \pi^2     \over 3 m_a^3  }  \beta' n_{\gamma c}^2  
       -   { 16 \pi^2      \over  3 m_a^3  }\beta' {   n_{\gamma c} n_{\gamma d} } ]   ~.
\end{flalign}\\

Counting time in units of axion decay constant $1/\Gamma_a$, then defining a new dimensionless variable $t_a=t\Gamma_a$ and measuring volume in unit of axion Compton volume ${16\pi^2 \over m_a^3}$, the system of rate equations can be expressed as
\begin{flalign} \label{main-gc}
&    {d n_{\gamma c}^{\text C}  \over dt_a} 
=  [ 2 n_{ac}^{\text C}     +     {   1  \over   \beta' } n_{ac}^{\text C} n_{\gamma c}^{\text C}   -    {  1     \over 3    }  (\beta'+{3\over2})  (n_{\gamma c}^{\text C})^2  ]    - {\Gamma_s \over \Gamma_a} n_{\gamma c}^{\text C}  ~.
\end{flalign}

\begin{flalign} \label{main-ac}
&    {d n_{ac}^{\text C}  \over dt_a}  
=    [ - n_{ac}^{\text C}     -     {1   \over 2 \beta' } n_{ac}^{\text C} n_{\gamma c}^{\text C}  +    { \beta'   \over 6  }   (n_{\gamma c}^{\text C})^2 ]    ~.
\end{flalign}

\begin{flalign} \label{main-gd}
    {d n_{\gamma d}^{\text C}  \over dt_a}
= &    2 n_{ac}^{\text C}   +  2  n_{ad}^{\text C}  +     { 1 \over \beta' } ( n_{ac}^{\text C} n_{\gamma c}^{\text C} +      n_{ac}^{\text C}  n_{\gamma d}^{\text C}  +     n_{\gamma c}^{\text C}  n_{ad}^{\text C} )\\ \nonumber
&     -  { 1  \over 3  } (\beta' + {3\over2}) (n_{\gamma c}^{\text C}) ^2  
       -   {  2      \over  3   } (\beta' + {3\over2}) {   n_{\gamma c}^{\text C}  n_{\gamma d}^{\text C}  }    - {\Gamma_s \over \Gamma_a} n_{\gamma d}^{\text C}  ~.
\end{flalign}

\begin{flalign} \label{main-ad}
    {d n_{ad}^{\text C}  \over dt_a}  
= &   - n_{ac}^{\text C}   -    n_{ad}^{\text C}  -     { 1  \over 2 \beta' } ( n_{ac}^{\text C} n_{\gamma c}^{\text C} +      n_{ac}^{\text C} n_{\gamma d}^{\text C}  +     n_{\gamma c}^{\text C}  n_{ad}^{\text C}  )\\ \nonumber
&     +  { \beta'   \over 6  }   (n_{\gamma c}^{\text C}) ^2  
       +   { \beta'   \over  3    } {   n_{\gamma c}^{\text C}  n_{\gamma d}^{\text C}  }   ~.
\end{flalign}

Equations \eqref{main-gc} \eqref{main-ac} \eqref{main-gd} \eqref{main-ad} are our main results. They can be applied in a variety of circumstances and we will show one specific example.

\section{Example: lasing axions clustered near a solar mass host object}

The following example is meant to be illustrative but the scenario described in this example is not necessarily physically. Non the less it will help place our results in context.

We have numerically solved the above system of rate equations for the case of a one solar mass, $M=M_{\odot}$ Schwarzschild host. We assume there is a hadronic axion ($\sim$ 3 eV) cluster (Hadronic axions are not currently favored.) of diameter $600$ m with  density 900 kg/m$^3$ (roughly the density of water ice) which is equivalent to an initial axion number density of about $7.56\times10^{18}$ times the unit axion Compton number density (Compton number density means that there is about 1 axion inside 1 Compton volume defined by the Compton wavelength of the axion). If the cluster is placed at 40 AU (this corresponds to the radius of the Kuiper belt in the solar system) from the host, the relativity index becomes $\beta'=2.156\times10^{-10}$. The uniform photon density $n_{\gamma c}$ grows exponentially on a time scale of $10^{-28} /\Gamma_a$.
See Figures 1.\\

\begin{figure}[ht]
\includegraphics[scale=0.7]{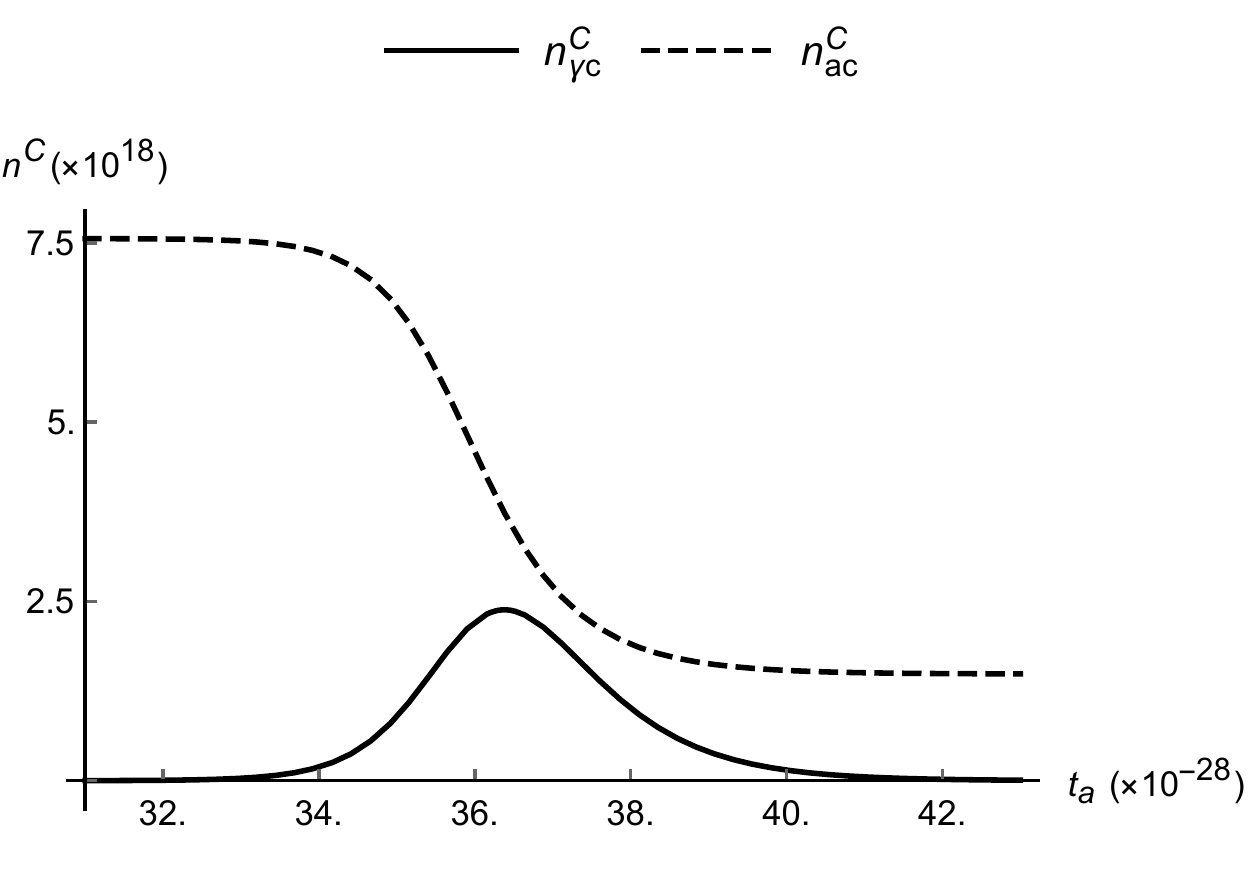}
\caption{ The uniform axion density $n_{ac}$ decreases exponentially on the same temporal scale as the photos.}
\label{fig:WormholeEmbed}
\end{figure}

\begin{figure}[ht]
\includegraphics[scale=0.7]{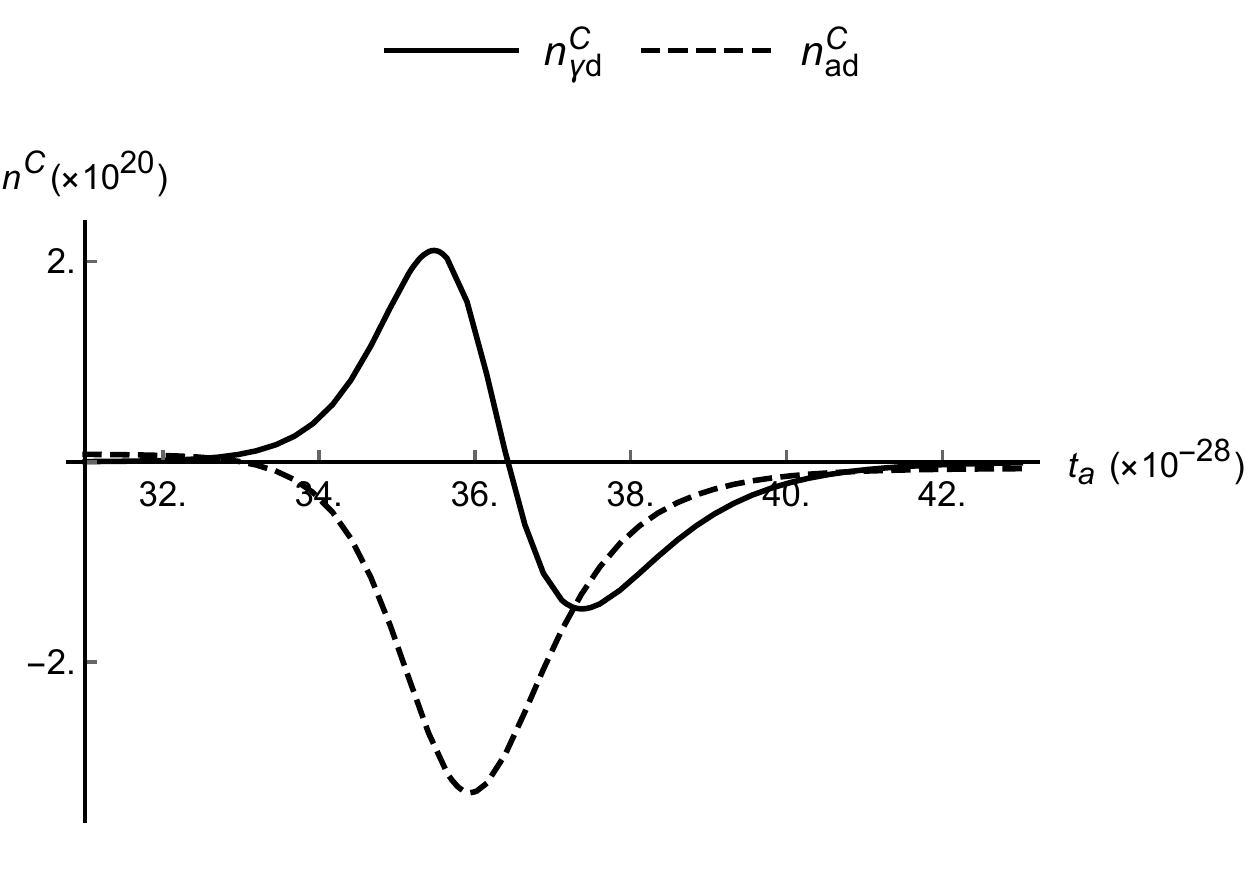}
\caption{The distorted photon density formed a sharp pulse. The distorted axion density also formed a sharp pulse with the amplitude being the opposite of that of distorted photon. }
\label{fig:WormholeEmbed}
\end{figure}

\begin{figure}[ht]
\includegraphics[scale=0.7]{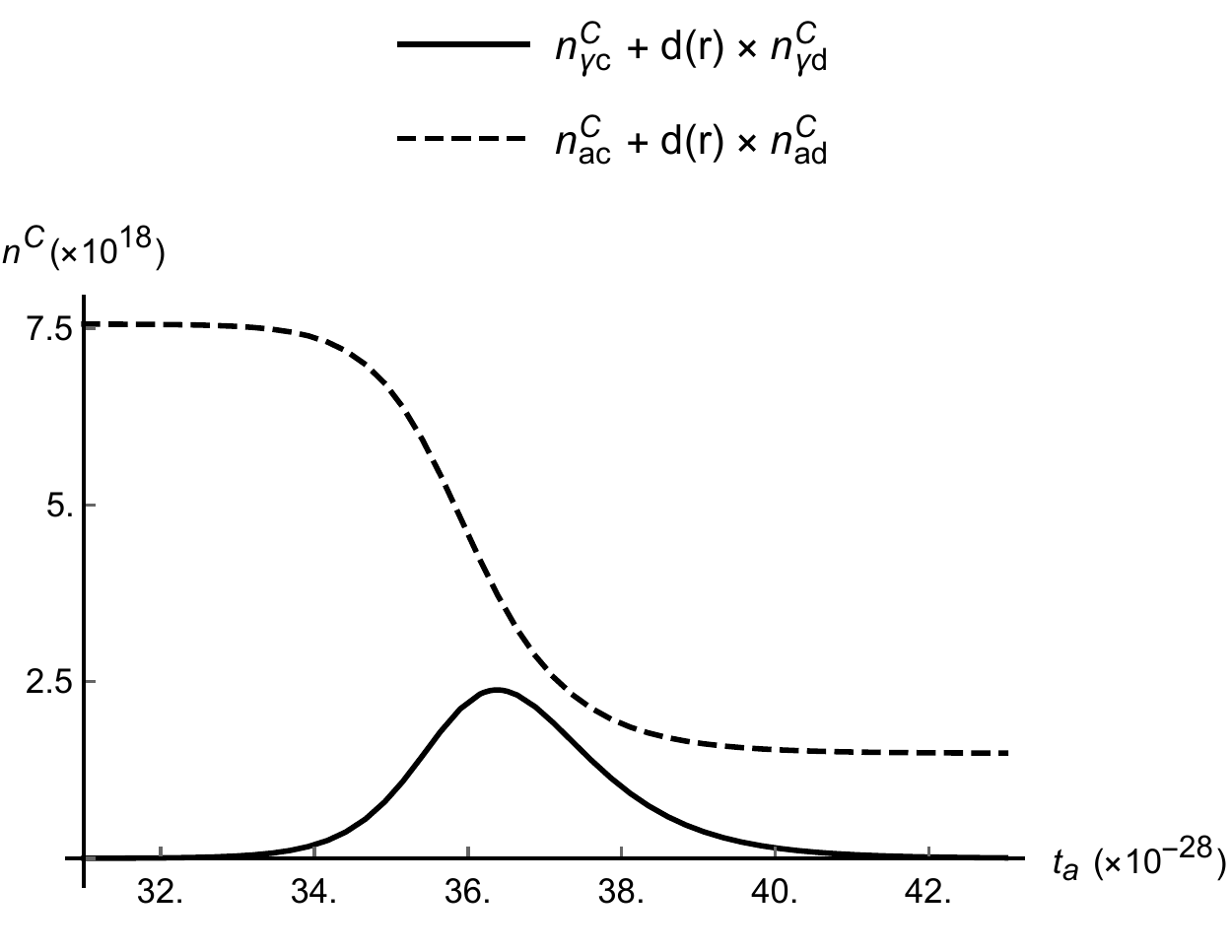}
\caption{The total particle densities}
\label{fig:WormholeEmbed}
\end{figure}
Since the distortion factor $d(r)\sim-2.465\times10^{-10}$ is very small (see Figure 2.), the total photon and axion number density are  affected very little. (See Figure 3.) The detailed photon radiation outcome, such as growth time and pulse height, are highly dependent on the initial axion number density, as the surface loss would affect low density axion clusters more noticeably than high density ones.\\

For the numerical calculation we selected a section of the axion cluster that is a section of a cone
of height $600$ m, i.e.,  a frustum, which we get by including the factors    $\Theta(\theta)\Theta(10^{-8}-\theta)\Theta(\phi)\Theta(10^{-8}-\phi)$ to constrain the axion and photon occupation numbers \eqref{axion-occu} and \eqref{photon-occu}, since $g_{00}$ in equation \eqref{DnPhotonDt} does not depend on the $\theta$ or $\phi$ if the host is of Schwarzschild or Reissner-Nordstr{\"o}m type. Including these factors  makes our results applicable to asteroid size axion cluster. These factors will not change the $\beta'$. During the lasing time scale, axions move only a few micrometer if the previous $\beta'$ is  the maximum velocity of the axions. This means that the axions would  be confined in the region described by these $\Theta$ factors during lasing. Surface loss terms may need to be changed here, therefore it is not guaranteed that every point in the cluster would lase as the figures suggest, but some part of the cluster will. Other examples are easily handled by this approach, for instance, ring type axion clusters are also able to be described by our results by adding similar angular factors and changing surface loss terms.

 \newpage

\section{Discussion and Conclusion}

We provided an example that shows lasing is possible for axions far away from the host object. Our treatment of gravity is different from references mentioned in the introduction. However, this  leads to a slightly nuanced understanding. Self gravity is considered in studies on axion cluster such as \cite{Braaten:2018lmj}. Self gravity is also considered in \cite{Kephart:1994uy} by using Newtonian gravity. What we present here is the lasing of axion cloud bounded by a host object, not a self bounded system. The parameters  in the example need to be somewhat carefully chosen, for lasing to occur., and this seems to imply that axion cloud lasing requires stricter conditions than axion cluster lasing. An axion cloud that requires a host object to bind typically has a lower density which makes lasing more difficult. The evolution equation suggests that low value of the maximum momentum $\beta'$ amplifies lasing the process. The advantage for lasing of self bound axion clusters is that the cluster can form a Bose-Einstein condensate which makes $\beta'$ essentially 0 if the coordinate system is judiciously chosen. Therefore, we conclude that self bound axion clusters may have a better chance of lasing than axion cloud bounded by a host.\\

One may argue that an axion cluster near a massive host would be freely falling and thus, by the equivalence principle in the cluster inertial frame, the usual physics applies and there is no need for any gravitational corrections. This is not the case since the cluster has finite size and there are tidal forces, differential rotation, etc to consider, plus frame dragging etc. in Kerr spacetime. The equation for lasing does not depend on the choice of coordinates, only on whether it is put into a covariant form and that is what we did. But a different observer finds different evolution equation if a specific set of coordinate is chosen. More importantly, the observer need not be in the axion's inertial frame. \\

Axions are copiously produced at the QCD phase transition. A possible way to detect cosmological axions is through the lasing of axion clusters. If axions are a component of the cold dark matter, then their density perturbations in the early universe can grow to form highly over dense regions. If the density is high enough, then ambient photons from the CBM or from axion spontaneous decay can cause these axion clumps to begin to undergo stimulated emission, i.e., to lase   \cite{Kephart:1986vc,Tkachev:1987cd,Kephart:1994uy}.\\

Besides the initial density perturbations, other axion structures can form from the evolution of those perturbations, e.g., caustics from the  infall of galactic axions    \cite{Sikivie:1997ng,Duffy:2008dk}. Yet another possibility is that axions can be produced via a Penrose type process in the neighborhood of primordial black holes (PBHs).\\

Such black holes could be another result of early universe density perturbations which may come from axions, but could also have other origins. If the PBHs have
angular momentum, which would be most likely but not necessarily due to mergers, and if their masses are in the right range $M_{BH} \sim \lambda_C^{-1}$, where $ \lambda_C$ is the Compton wave length of the axion, then  superradience can lead to axion in an $n,\ell,m = 2,1,1$ hydrogenlike orbit around these primordial Kerr black holes. Again, if the axion density is high enough, then lasing can commence. Such objects have been proposed as the source of fast radio bursts (FRBs) \cite{Rosa:2017ury}. Other axion lasing mechanisms for FRBs have  been suggested in \cite{Levkov:2020txo,Buckley:2020fmh}. \\

The results given here will allow us to improve on a broad class of  models of   lasing   axion clusters in   static 
spacetimes. The results also hold for the
nonstatic case when the lasing happens so fast that the static approximation holds. The improvements  
include both special and general relativistic corrections. We plan to study specific physically relevant  examples in future work.

\newpage
\vspace{0.2cm}

\section{Appendix: Some calculational details}
In this appendix we give some of the calculational details that were avoided in the text in order to allow the discussion to proceed more naturally. 

\subsection{$dX_\text{LIMS}$}
We recall, in more than the usual amount of detail, how the Lorentz invariant phase space originates in Minkowski space in order to generalize it to curved space.  Specifically, the one body Lorentz invariant momentum space measure is
\begin{flalign} \nonumber
\int dX_\text{LIMS} = & \int {d^3p\over (2\pi)^3 2p^0} ~.
\end{flalign}
 This comes about in the following way.
In Minkowski spacetime, the quantity $dp^0 dp^1 dp^2 dp^3$ which equals to $\sqrt{|\eta|} dp^0 dp^1 dp^2 dp^3 $ is Lorentz invariant, where $\eta$ is the determinant of the Minkowski metric. This is obvious because it measures the 4-momentum volume in any Lorentz coordinate system. The form of $dp^0 dp^1 dp^2 dp^3$ is preserved under Lorentz transformation.\\

For a particle in special relativity, when we count how many possible momentum states it possesses, $p^0$, $p^1$, $p^2$, $p^3$ can not just take any values. These values have to satisfy the normalization condition $p^\mu p_\mu = m^2$. The first component  is just the energy of the particle, so in addition, we need to require that $p^0>0$. Therefore, a new quantity that is both Lorentz invariant and counts the number of all possible momentum states is
\begin{flalign} \nonumber
& \int_{p_0 = -\infty}^{p_0 = +\infty} dp^0 dp^1 dp^2 dp^3 \delta( p^\mu p_\mu - m^2 ) \Theta(p^0)   \\ \nonumber
= &    { d^3p \over 2\sqrt {\vec{p}^2 + m^2} } = { d^3p \over 2 p^0  }  = (2\pi)^3 dX_\text{LIMS}  ~.
\end{flalign}

The quantity
$\sqrt{|\eta|} dx^0 dx^1 dx^2 dx^3 = dx^0 dx^1 dx^2 dx^3 $ is a Lorentz invariant pseudoscalar. The small change $d{(\frac{\tau}{m})}$ is also a Lorentz invariant, where $\tau$ and $m$ are the proper time and the rest mass of the particle, respectively. Hence we have another Lorentz invariant quantity
\begin{flalign} \nonumber
& m{ dx^0 \over d\tau } dx^1 dx^2 dx^3  = p^0 d^3 r  ~.
\end{flalign}
Finally, from the product of two pseudoscalars $ dx^0 dx^1 dx^2 dx^3 $ and $dp^0 dp^1 dp^2 dp^3$, which is a scalar, the Lorentz invariant phase space measure arises naturally,
\begin{flalign} \nonumber
& { 2 m \over d\tau } dx^0  dx^1 dx^2 dx^3 \int dp^0 dp^1 dp^2 dp^3   \delta( p^\mu p_\mu - m^2 ) \Theta(p^0) \\\nonumber
=& 2   p^0 d^3 r   { d^3p \over 2 p^0  } =   d^3 r d^3p  =  dX_\text{LIPS}  ~.
\end{flalign}
This is the quantity that we will generalize to curved space below.

\subsection{Occupation number and number density of particles in Minkowski spacetime}
The infinitesimal line element in Minkowski metric using polar coordinates reads
\begin{flalign} \label{Min-metric}
ds^2=- dt^2 +  dr^2 + r^2d\theta^2 +  r^2\sin^2\theta d\phi^2~.
\end{flalign}
In these coordinates the particle number density is given by
\begin{flalign} \nonumber
n(\vec{r},t)=&\int f(\vec{p},\vec{r}, t) { d^3p \over (2\pi)^3 }=\int  {f(\vec{p},\vec{r}, t) \over (2\pi)^3 }  dp^x dp^y dp^z \\ \nonumber
=& \int  {f(\vec{p},\vec{r}, t) \over (2\pi)^3 } { \partial(p^x , p^y , p^z) \over \partial( p^r , p^\theta , p^\phi ) }  dp^r dp^\theta dp^\phi  \\ \nonumber
=& \int  {f(\vec{p},\vec{r}, t) \over (2\pi)^3 } r^2 \sin\theta  dp^r dp^\theta dp^\phi \\ \nonumber
=& \int  {f(\vec{p},\vec{r}, t) \over (2\pi)^3 } \sqrt{|\eta_\text{\tiny E}|} dp^r dp^\theta dp^\phi ~,
\end{flalign}
where $f$ is the occupation number and $\eta_\text{\tiny E}$ is the determinant of the 3 dimensional Euclidean metric in spherical coordinates. The total number of the particle is
\begin{flalign} \nonumber
N(t) =\int n(\vec{r}, t)  r^2 \sin\theta dr  d\theta d\phi =\int n(\vec{r}, t)  \sqrt{|\eta_\text{\tiny E}|}  dr  d\theta d\phi ~.
\end{flalign}

\subsection{Covariant measures in static spacetime}

For any species of particles, the occupation number $f(\vec{p},\vec{r},t)$ in Minkowski spacetime should change to $f(p^i,x^\alpha)$, where $p^i$ is the three spatial component of the 4-momentum of the particle and $x^\alpha$ is the spacetime coordinate of the particle. The occupation number does not depend on the first component $p^0$ of the 4-momentum of the particle since normalization of the 4-momentum still holds in Schwarzschild spacetime. The coordinate time $t$ is the time measured by a stationary observer at $r=\infty$ with 4-velocity $\xi = (1,0,0,0)$.\\

The concept of Lorentz transformation is not very useful in static spacetime because there is no global inertial frame. At a single event, the spacetime can be treated locally as flat and special relativity is still applicable, Lorentz transformation still provides the relationship between two nearby observers from two local inertial frames having relative velocity in that infinitesimal region. As soon as an observer, for example,  moves forward along the geodesic a finite distance, the previous established Lorentz transformation loses its meaning.\\

The curvature of Schwarzschild spacetime presented in the form \eqref{schw-metric} is from the perspective of a special observer  stationary  at $r=\infty$ who sets the origin $r=0$ as the location of the singularity. The advantage of the metric form \eqref{schw-metric} is that it displays the symmetry of the metric conspicuously. When we change coordinates, the curvature of spacetime would be shown from the perspective of other observers. So even if the functionality of Lorentz transformation between inertial frames is irrecoverable in curved spacetime, we can embrace the general coordinate transformation under which general relativity demonstrates general covariance. For example, in Schwarzschild spacetime, the coordinates can be expressed as Schwarzschild coordinates, or Kruskal-Szekeres coordinates, or other equivalent transformations, but the general covariant quantities are valid without referring to the specific coordinate system.\\

 In curved spacetime, when we count how many possible momentum states a particle can have, $p^0$, $p^1$, $p^2$, $p^3$ are not arbitrary. They still have to satisfy a normalization condition $g_{\mu\nu} p^\mu p^\nu = - m^2$ as they did in Minkowski space. The minus sign is due to the signature convention we chose for the curved space time metric. The first component is still related to the energy of the particle. A stationary particle has 4-momentum $p^\mu=(p^0,0,0,0)$, and its 4-velocity is $v^\mu = (p^0/m,0,0,0)$. A co-stationary observer at the location of the particle would measure the energy of the particle to be $m=-g_{\mu\nu} p^\mu v^\nu = -g_{00}{(p^0)^2\over m}$. $p^0 = \pm { m \over \sqrt{-g_{00}} }$, where we choose $p^0$ such that it reduces to $m$ in the flat spacetime limit. So in addition, we may as well require that $p^0>0$. Moreover, $p^0>0$ implies that ${dt\over d\tau} > 0$, or the proper time and coordinate time flow in the same direction. Similar to the Lorentz invariant measure $dX_\text{LIMS}$, now there is a general covariant measure $dX_\text{GCMS}$,
\begin{flalign} \nonumber
dX_\text{GCMS} =& \int \sqrt{|g|} dp^0 dp^1 dp^2 dp^3 \delta(  g_{\mu\nu} p^\mu p^\nu + m^2) \Theta(p^0)   ~.
\end{flalign}
The general covariant 4-momentum volume element is $\sqrt{|g|} dp^0 dp^1 dp^2 dp^3$, regardless of the metric or the coordinates that give a specific form to the metric, where $g$ is the determinant of the metric.
This abstract form of momentum space is as far as we can go without knowing of anything specific of the spacetime. It is applicable to not only Schwarzschild spacetime, but to any stationary spacetime, including the Kerr spacetime.\\

Now let us consider static spacetime where $g_{0\mu}=g_{00}\delta_{0\mu}$, which is equivalent to the choice of synchronous gauge. In this case we can have a static covariant measure $dX_\text{SCMS}$ which is a 4-momentum volume measure that is compatible with all the static spacetime and their coordinates transformations that keep $g_{0\mu}=g_{00}\delta_{0\mu}$. 
\begin{flalign} \nonumber
&dX_\text{SCMS} \\ \nonumber
= & \sqrt{|g|} \int_{p_0 = 0}^{ +\infty} dp^0 dp^1 dp^2 dp^3 \delta[   g_{00} (p^0)^2 + g_{ij} p^i p^j + m^2 ] ~.
\end{flalign}
This can be rewritten as
\begin{flalign} \nonumber
&dX_\text{SCMS} \\ \nonumber
= &  \sqrt{|g|} dp^1 dp^2 dp^3 \int_{ 0}^{+\infty} dp^0 \\ \nonumber
                              &  \times \delta[g_{00} (p^0 - \sqrt { { g_{ij} p^i p^j + m^2 \over -g_{00} }}) (p^0 + \sqrt { { g_{ij} p^i p^j + m^2 \over -g_{00} }})  ]  ~.
\end{flalign}
Using the properties of the Dirac $\delta$-function, we obtain
\begin{flalign} \nonumber
&dX_\text{SCMS} \\ \nonumber
= &  \sqrt{|g|} dp^1 dp^2 dp^3 \int_{ 0}^{+\infty} dp^0 \\ \nonumber
                              &  \times [ {1\over - 2 g_{00} \sqrt { { g_{ij} p^i p^j + m^2 \over -g_{00} }} } \delta(p^0 - \sqrt { { g_{ij} p^i p^j + m^2 \over -g_{00} }})   \\ \nonumber
                              &+ {1\over -2 g_{00} \sqrt { { g_{ij} p^i p^j + m^2 \over -g_{00} }} } \delta(p^0 + \sqrt { { g_{ij} p^i p^j + m^2 \over -g_{00} }})   ] . 
\end{flalign}
Carrying out the integral leads to the final form
\begin{flalign} \nonumber
&dX_\text{SCMS} \\ \nonumber
= &    { \sqrt{|g|} dp^1 dp^2 dp^3 \over 2 \sqrt{-g_{00}} \sqrt {  g_{ij} p^i p^j + m^2  }   }   
=   { \sqrt{ -g_{00} ||g_{ij}||} dp^1 dp^2 dp^3 \over 2 \sqrt{-g_{00}} \sqrt { - g_{00} (p^0)^2  }   }  \\ \label{scms}
=& { \sqrt{   ||g_{ij}||} dp^1 dp^2 dp^3 \over 2   \sqrt { - g_{00}    } p^0  } ~,
\end{flalign}
where $|g_{ij}|$ is the determinant of the metric at the 3-surface $dx^0=0$ in the static spacetime of
$ds^2=g_{00}(dx^0)^2 + g_{ij}dx^i dx^j ~.$

\subsection{Integration in $k_1$ space}
We integrate the $k_1^i$ part of \eqref{evolution} first,
\begin{flalign} \nonumber
&  2k^0 \frac{d f_\lambda ( k^i )  }{dt} \\ \nonumber
=& \frac{4 m_{a} \Gamma_{a} }{ \pi }    \int  (  g_{ij} { p^i p^j  \over m_a^2 } +  1  )^{-1/2} 
      { \sqrt{   ||g_{ij}||} dp^1 dp^2 dp^3 \over   2   \sqrt { - g_{00}    } p^0  }  \\ \nonumber
&\times \int { \sqrt{   ||g_{ij}||} dk_1^1 dk_1^2 dk_1^3 \over   2   \sqrt { - g_{00}    } k_1^0  }
       { 1 \over \sqrt{|g|} }    \delta (p^0-k^0-k_1^0)   \\ \nonumber
&\times \delta (p^1-k^1-k_1^1) \delta (p^2 - k^2 - k_1^2) \delta (p^3 - k^3 - k_1^3 ) \\ \nonumber
&\times \{  f_a  ( p^i ) [ 1+ f_\lambda ( k^i ) + f_\lambda ( k_1^i ) ] - f_\lambda ( k^i )f_\lambda ( k_1^i ) \}.
\end{flalign}
The only function that depends on $k_1^1$, $k_1^2$ and $k_1^3$ is $f_\lambda ( k_1^i ) = f_\lambda ( k_1^1, k_1^2, k_1^3 )$. Therefore, after the three $\delta$ function are integrated, $f_\lambda ( k_1^i )$ will change to $ f_\lambda ( p^1-k^1 , p^2-k^2, p^3-k^3 ) = f_\lambda ( p^i-k^i ) $.\\

$k_1^0$ is an implicit function of $k_1^1$, $k_1^2$ and $k_1^3$ as given by the photon energy momentum relation \eqref{photon-e-p}. After integration $k_1^0$ would change to
\begin{flalign} \nonumber
k_1^0 \quad\rightarrow\quad       \sqrt{ g_{ij}  (p^i-k^i) (p^j-k^j) \over  - g_{00} }   ~,
\end{flalign}
where $g_{ij}  (p^i-k^i) (p^j-k^j)$ is the square of the magnitude of the 3-momentum $p^i-k^i$. Given that  $ g = - g_{00}  |g_{ij}| $, we now have
\begin{flalign} \nonumber
&  2k^0 \frac{d f_\lambda ( k^i )  }{dt} \\ \nonumber
=& \frac{4 m_{a} \Gamma_{a} }{ \pi }    \int  (  g_{ij} { p^i p^j  \over m_a^2 } +  1  )^{-1/2} 
      { \sqrt{   ||g_{ij}||} dp^1 dp^2 dp^3 \over   2   \sqrt { - g_{00}    } p^0  }  \\ \nonumber
&\times   {   1 \over   2    \sqrt{ g_{ij}  (p^i-k^i) (p^j-k^j)  }  }
       { 1 \over \sqrt{  - g_{00} } }      \\ \nonumber
&\times \delta [  p^0 - k^0- \sqrt{ g_{ij}  (p^i-k^i) (p^j-k^j) \over  - g_{00} }  ]  \\ \nonumber
&\times \{   f_a  ( p^i ) [ 1+ f_\lambda ( k^i ) + f_\lambda ( p^i-k^i ) ] - f_\lambda ( k^i )f_\lambda ( p^i-k^i ) \}.
\end{flalign}

From equation \eqref{mass-e-p}, $ g_{ij}   p^i p^j$ can be rewritten in terms of $p^0$,
\begin{flalign}   \nonumber
(  g_{ij} { p^i p^j  \over m_a^2 } +  1  )^{-1/2} = &   { m_a \over \sqrt{-g_{00}} p^0 } ~.
\end{flalign}
This leads to
\begin{flalign} \nonumber
&  2k^0 \frac{d f_\lambda ( k^i )  }{dt} \\ \nonumber
=& \frac{4 m_{a} \Gamma_{a} }{ \pi }    \int  { m_a \over \sqrt{-g_{00}} p^0 }
      { \sqrt{   ||g_{ij}||} dp^1 dp^2 dp^3 \over   2   \sqrt { - g_{00}    } p^0  }  \\ \nonumber
&\times   {   1 \over   2    \sqrt{ g_{ij}  (p^i-k^i) (p^j-k^j)  }  }
       { 1 \over \sqrt{  - g_{00} } }      \\ \nonumber
&\times \delta [  p^0 - k^0- \sqrt{ g_{ij}  (p^i-k^i) (p^j-k^j) \over  - g_{00} }  ]  \\ \nonumber
&\times \{   f_a  ( p^i ) [ 1+ f_\lambda ( k^i ) + f_\lambda ( p^i-k^i ) ] - f_\lambda ( k^i )f_\lambda ( p^i-k^i ) \} ~.
\end{flalign}
Canceling common factors, we arrive at \eqref{rate-equa-1}
\begin{flalign} \nonumber
&  2k^0 \frac{d f_\lambda ( k^i )  }{dt} \\ \nonumber
=& \frac{4 m_a \Gamma_{a} }{ \pi  ( - g_{00} )  }    \int  {m_a\over p^0 } { \sqrt{   ||g_{ij}||} dp^1 dp^2 dp^3 \over   2  \sqrt{  - g_{00} } p^0   }  \\ \nonumber
&\times  {   1 \over   2    \sqrt{ g_{ij}  (p^i-k^i) (p^j-k^j)  }  }  \\ \nonumber
&\times \delta [  p^0 - k^0- \sqrt{ g_{ij}  (p^i-k^i) (p^j-k^j) \over  - g_{00} }  ]  \\  \nonumber
&\times \{   f_a  ( p^i ) [ 1+ f_\lambda ( k^i ) + f_\lambda ( p^i-k^i ) ] - f_\lambda ( k^i )f_\lambda ( p^i-k^i ) \} ~.
\end{flalign}

\subsection{Integration in $p^i$ space with isotropic occupation number and Riemann normal coordinates}
Here, we can proceed to integrate if occupation numbers are assumed to be isotropic. Occupation numbers depend on the 3-momentum only through its norm,
\begin{flalign} \nonumber
f_a  ( p^i ) =& f_a  ( \sqrt{ g_{ij} p^i p^j } )  \\\nonumber
f_\lambda ( k^i ) =& f_\lambda ( \sqrt{ g_{ij} k^i k^j } ) ~.
\end{flalign}
Applying the isotropic assumption to  equation \eqref{rate-equa-1} results in
\begin{flalign} \label{rate-equa-3}
&  2 k^0 \frac{d f_\lambda ( \sqrt{ g_{ij} k^i k^j } )  }{dt} \\ \nonumber
=& \frac{4 m_a \Gamma_{a} }{ \pi  ( - g_{00} )  }    \int  {m_a\over p^0 } { \sqrt{   ||g_{ij}||} dp^1 dp^2 dp^3 \over   2  \sqrt{  - g_{00} } p^0   }  \\ \nonumber
&\times  {   1 \over   2    \sqrt{ g_{ij}  (p^i-k^i) (p^j-k^j)  }  }  \\ \nonumber
&\times \delta [  p^0 - k^0- \sqrt{ g_{ij}  (p^i-k^i) (p^j-k^j) \over  - g_{00} }  ]  \\ \nonumber
&\times \{   f_a  ( \sqrt{ g_{ij} p^i p^j } )   [ 1+ f_\lambda ( \sqrt{ g_{ij} k^i k^j } )  \\ \nonumber
& + f_\lambda \left( \sqrt{ g_{ij} ( p^i - k^i )( p^j -  k^j ) } \right)  ]  \\ \nonumber
& - f_\lambda ( \sqrt{ g_{ij} k^i k^j } )   f_\lambda \left( \sqrt{ g_{ij} ( p^i - k^i )( p^j -  k^j ) } \right)  \} ~.
\end{flalign}

Equation \eqref{rate-equa-1} is valid at event $\mathcal P_0(t_{\mathcal P_0}, x^i_{\mathcal P_0})$.
On the 3-surface $ds^2= g_{ij}dx^i dx^j$, at point $\mathcal P_0$ we employ Riemann normal coordiantes $x_\text{\tiny R}^i$ on an infinitesimal small patch around $\mathcal P_0$.
We write  the metric on the infinitesimal 3-patch in these coordinates as
\begin{flalign} \nonumber
ds^2= {\mathcal G}_{ij}(x_\text{\tiny R}) dx_\text{\tiny R}^i dx_\text{\tiny R}^j ~.
\end{flalign}
None of the temporal components are  subject to change in  equation \eqref{rate-equa-3}, because the Riemann normal coordiantes are employed on the 3-surface. But the coordinate labels have to be changed from $(t, x^i)$ to $\mathcal P_0$, to remind us that we are focusing on the physics only at $\mathcal P_0$. Hence we will make the following replacements 
\begin{flalign} \nonumber
k^0 (t, x^1, x^2, x^3)         &\quad\Rightarrow\quad        k^0 \Big|_{\mathcal P_0}   \\ \nonumber
p^0 (t, x^1, x^2, x^3)         &\quad\Rightarrow\quad        p^0 \Big|_{\mathcal P_0}   \\ \nonumber
g_{00} (t, x^1, x^2, x^3)         &\quad\Rightarrow\quad        g_{00} \Big|_{\mathcal P_0}   ~.
\end{flalign}

If there is no isotropic assumption, the occupation numbers would depend on each momentum spatial component $p^i$. But in general we do not know how to express $p^i$ under generic Riemann normal coordinates. It would be some generic function $p^i=F(x_\text{\tiny R}, p_\text{\tiny R})$, where $p_\text{\tiny R}$ is the corresponding momentum in Riemann normal coordinates. Calculations are difficult to proceed without isotropic assumption. \\

The isotropic assumption  is required so that the spatial quantites in equation \eqref{rate-equa-3} can be written in general covariant form (general covariant only on the 3-surface). General covariant quantities keep their form when switching to Riemann normal coordinates $x_\text{\tiny R}^i$,
\begin{flalign} \nonumber
\sqrt{   ||g_{ij}||} dp^1 dp^2 dp^3      
\,\Rightarrow\,&   \sqrt{ ||  {\mathcal G}_{ij}   ||} dp_\text{\tiny R}^1  dp_\text{\tiny R}^2 dp_\text{\tiny R}^3 \Big|_{\mathcal P_0} \\\nonumber
 \sqrt{ g_{ij}  (p^i-k^i) (p^j-k^j)  }        
\,\Rightarrow\,&      \sqrt{  {\mathcal G}_{ij} ( p_\text{\tiny R}^i   -k_\text{\tiny R}^i  )( p_\text{\tiny R}^j   -k_\text{\tiny R}^j  )  }\Big|_{\mathcal P_0} 
\end{flalign}
\begin{flalign} \nonumber
\sqrt{ g_{ij} p^i p^j }   &\, \Rightarrow\,     \sqrt{   {\mathcal G}_{ij}    p_\text{\tiny R}^i    p_\text{\tiny R}^j  } \Big|_{\mathcal P_0}   \\\nonumber
\sqrt{ g_{ij} k^i k^j }   &\, \Rightarrow\,     \sqrt{   {\mathcal G}_{ij}    k_\text{\tiny R}^i    k_\text{\tiny R}^j  } \Big|_{\mathcal P_0} ~.
\end{flalign}
With these substitutions equation \eqref{rate-equa-3} becomes
\begin{flalign} \label{rate-equa-4}
&  2 k^0 \Big|_{\mathcal P_0}  \frac{d f_\lambda ( \sqrt{ {\mathcal G} k_\text{\tiny R}^i k_\text{\tiny R}^j } \Big|_{\mathcal P_0} , \mathcal P_0 )  }{dt} \\ \nonumber
=& \frac{4 m_a \Gamma_{a} }{ \pi  ( - g_{00} \Big|_{\mathcal P_0} )  }    \int  {m_a\over p^0  \Big|_{\mathcal P_0}  } 
{ \sqrt{ ||  {\mathcal G}_{ij}  ||} dp_\text{\tiny R}^1 dp_\text{\tiny R}^2 dp_\text{\tiny R}^3  \Big|_{\mathcal P_0}    \over   2  \sqrt{  - g_{00} \Big|_{\mathcal P_0}   }  p^0 \Big|_{\mathcal P_0}    }  \\ \nonumber
&\times  {   1 \over   2    \sqrt{ {\mathcal G}_{ij}  (p_\text{\tiny R}^i-k_\text{\tiny R}^i) (p_\text{\tiny R}^j-k_\text{\tiny R}^j)} \Big|_{\mathcal P_0} }  \\ \nonumber
&\times \delta [  p^0  \Big|_{\mathcal P_0}  - k^0  \Big|_{\mathcal P_0} - \sqrt{ {\mathcal G}_{ij}  (p_\text{\tiny R}^i-k_\text{\tiny R}^i) (p_\text{\tiny R}^j-k_\text{\tiny R}^j) \over  - g_{00} } \Big|_{\mathcal P_0}  ]  \\ \nonumber
&\times \{   f_a  ( \sqrt{ {\mathcal G}_{ij} p_\text{\tiny R}^i p_\text{\tiny R}^j }  \Big|_{\mathcal P_0} )   [ 1+ f_\lambda ( \sqrt{ {\mathcal G}_{ij} k_\text{\tiny R}^i k_\text{\tiny R}^j }  \Big|_{\mathcal P_0}  )  \\ \nonumber
& + f_\lambda\left( \sqrt{ {\mathcal G}_{ij} ( p_\text{\tiny R}^i - k_\text{\tiny R}^i )( p_\text{\tiny R}^j -  k_\text{\tiny R}^j ) }  \Big|_{\mathcal P_0}  \right)  ]  \\ \nonumber
& - f_\lambda( \sqrt{ {\mathcal G}_{ij} k_\text{\tiny R}^i k_\text{\tiny R}^j }   \Big|_{\mathcal P_0} )   
    f_\lambda \left( \sqrt{ {\mathcal G}_{ij} ( p_\text{\tiny R}^i - k_\text{\tiny R}^i )( p_\text{\tiny R}^j -  k_\text{\tiny R}^j ) } \Big|_{\mathcal P_0} \right)  \} ~.
\end{flalign}

The metric ${\mathcal G}_{ij}$ at point $\mathcal P_0$ in Riemann normal coordinates $x_\text{\tiny R}^i$ is ${\mathcal G}_{ij}\Big|_{\mathcal P_0}=(+1,+1,+1)$, which means at point $\mathcal P_0$ on the 3-surface, the space is Euclidean. So the momentum differential volume at point $\mathcal P_0$ can be expressed as
\begin{flalign} \nonumber
     \sqrt{ ||  {\mathcal G}_{ij}  ||} dp_\text{\tiny R}^1 dp_\text{\tiny R}^2 dp_\text{\tiny R}^3  \Big|_{\mathcal P_0} 
 = | \vec{p}_\text{\tiny R}| ^2 d| \vec{p}_\text{\tiny R}| d(-\cos\theta_\text{\tiny R})d(\phi_\text{\tiny R})  \Big|_{\mathcal P_0}  ~.
\end{flalign}
where $|p_\text{\tiny R}|$ is the magnitude of the 3-momentum, which is invariant under the coordinate change
\begin{flalign}  \nonumber
| \vec{p}_\text{\tiny R}|^2  \Big|_{\mathcal P_0} = {\mathcal G}_{ij} p_\text{\tiny R}^i p_\text{\tiny R}^j  \Big|_{\mathcal P_0}  = g_{ij} p^i p^j  \Big|_{\mathcal P_0} ~.
\end{flalign}
Rewriting other quantities using the Euclidean notation, gives
\begin{flalign}  \nonumber
\sqrt{ {\mathcal G}_{ij} p_\text{\tiny R}^i p_\text{\tiny R}^j }  \Big|_{\mathcal P_0} 
= & | \vec{p}_\text{\tiny R} |  \Big|_{\mathcal P_0}  \\\nonumber
\sqrt{ {\mathcal G}_{ij} k_\text{\tiny R}^i k_\text{\tiny R}^j }  \Big|_{\mathcal P_0} 
= & | \vec{k}_\text{\tiny R} |  \Big|_{\mathcal P_0}  \\\nonumber
\sqrt{ {\mathcal G}_{ij}  (p_\text{\tiny R}^i-k_\text{\tiny R}^i) (p_\text{\tiny R}^j-k_\text{\tiny R}^j)} \Big|_{\mathcal P_0} 
=&  | \vec{p}_\text{\tiny R}- \vec{k}_\text{\tiny R}|  \Big|_{\mathcal P_0} ~,
\end{flalign}
The law of cosines still hold at the event $\mathcal P_0$,
\begin{flalign}  \label{law-cos}
 | \vec{p}_\text{\tiny R}- \vec{k}_\text{\tiny R}| ^2 \Big|_{\mathcal P_0}   =  | \vec{p}_\text{\tiny R}| ^2 \Big|_{\mathcal P_0}    + | \vec{k}_\text{\tiny R} | ^2  \Big|_{\mathcal P_0}    - 2\cos\theta_\text{\tiny R} \Big|_{\mathcal P_0}  | \vec{p}_\text{\tiny R} |  \Big|_{\mathcal P_0}  | \vec{k}_\text{\tiny R} |  \Big|_{\mathcal P_0} ~.
\end{flalign}
The photon 3-momentum $\vec{k}_\text{\tiny R}$ is an independent variable in the 3-Euclidean space around the event $\mathcal P_0$. As far as the integration process is concerned, we have the freedom to choose that in this 3-Euclidean space, the angle formed by $\vec{p}_\text{\tiny R}$ and $\vec{k}_\text{\tiny R}$ is $\theta_\text{\tiny R}$, or in other words, $ \vec{k}_\text{\tiny R}= | \vec{k}_\text{\tiny R} | \vec{e}_z$.\\

Equation \eqref{law-cos} can be expressed as
\begin{flalign}  \nonumber
& (-g_{00}) (p^0 - k^0)^2 \Big|_{\mathcal P_0}   \\ \nonumber
= & [ (-g_{00}) (p^0)^2 - m_a^2 ] \Big|_{\mathcal P_0} + (-g_{00}) (k^0)^2 \Big|_{\mathcal P_0} \\ \nonumber
- &   2\cos\theta_\text{\tiny R} \sqrt{ (-g_{00}) (p^0)^2 - m_a^2 }\sqrt{ (-g_{00}) (k^0)^2 } \Big|_{\mathcal P_0}~.
\end{flalign}
$\cos^2 \theta_\text{\tiny R}  \Big|_{\mathcal P_0} \leq1$ requires that
\begin{flalign}  \nonumber
   p^0   \Big|_{\mathcal P_0}    \geq     k^0  \Big|_{\mathcal P_0} + { m_a^2 \over 4  k^0  (-g_{00}) }   \Big|_{\mathcal P_0}  ~.
\end{flalign}
or 
\begin{flalign}  \nonumber
  k^0_\text{min}  \Big|_{\mathcal P_0}   \leq   k^0  \Big|_{\mathcal P_0}  \leq   k^0_\text{max}   \Big|_{\mathcal P_0} ~,
\end{flalign}
where
\begin{flalign}  \nonumber
 k^0_\text{max/min}  \Big|_{\mathcal P_0}  =& { 1 \over  2 } [\  p^0 \pm \sqrt{ (  p^0)^2 -  { m_a^2 \over -g_{00} }  } \ ] 
 \Big|_{\mathcal P_0}  \\ \nonumber
=&   { 1   \over  2 }  ( p^0 \pm  \sqrt{g_{ij} p^i p^j \over -g_{00} } )  \Big|_{\mathcal P_0}  ~.
\end{flalign}
This equation and equation \eqref{law-cos} provide a relationship between momentum of axions and  photons, all located at a single point where Riemann normal coordinates were applied. The transition from ${\mathcal P_0}$ to other events is straightforward because the bounds are written in covariant form.

With these notation replacements, \eqref{rate-equa-4} becomes
\begin{flalign} \nonumber
&  2 k^0 \Big|_{\mathcal P_0}  \frac{d f_\lambda ( | \vec{k}_\text{\tiny R} |  \Big|_{\mathcal P_0} , \mathcal P_0 )  }{dt} \\ \nonumber
=& \frac{4 m_a \Gamma_{a} }{ \pi  ( - g_{00} \Big|_{\mathcal P_0} )  }    \int  {m_a\over p^0  \Big|_{\mathcal P_0}  } 
{ | \vec{p}_\text{\tiny R}| ^2 d| \vec{p}_\text{\tiny R}| d(-\cos\theta_\text{\tiny R})d(\phi_\text{\tiny R})  \Big|_{\mathcal P_0}    \over   2  \sqrt{  - g_{00} \Big|_{\mathcal P_0}   }  p^0 \Big|_{\mathcal P_0}    }  \\ \nonumber
&\times  {   1 \over   2    | \vec{p}_\text{\tiny R}- \vec{k}_\text{\tiny R}|  \Big|_{\mathcal P_0} }  
\times \delta (  p^0  \Big|_{\mathcal P_0}  - k^0  \Big|_{\mathcal P_0} - { | \vec{p}_\text{\tiny R}- \vec{k}_\text{\tiny R}| \over  \sqrt{ - g_{00} } } \Big|_{\mathcal P_0} )  \\ \nonumber
&\times \{   f_a  ( | \vec{p}_\text{\tiny R} |  \Big|_{\mathcal P_0} )   [ 1+ f_\lambda ( | \vec{k}_\text{\tiny R} |  \Big|_{\mathcal P_0}  )  
   + f_\lambda ( | \vec{p}_\text{\tiny R}- \vec{k}_\text{\tiny R}| \Big|_{\mathcal P_0}  )  ]  \\ \nonumber
& - f_\lambda ( | \vec{k}_\text{\tiny R} |   \Big|_{\mathcal P_0} )   f_\lambda ( | \vec{p}_\text{\tiny R}- \vec{k}_\text{\tiny R}| \Big|_{\mathcal P_0} )  \} ~.
\end{flalign}
$\phi_\text{\tiny R} \Big|_{\mathcal P_0}$ is directly integrated to yield
\begin{flalign} \nonumber
&  2 k^0 \Big|_{\mathcal P_0}  \frac{d f_\lambda ( | \vec{k}_\text{\tiny R} |  \Big|_{\mathcal P_0} , \mathcal P_0 )  }{dt} \\ \nonumber
=& \frac{4 m_a \Gamma_{a} }{   ( - g_{00} \Big|_{\mathcal P_0} )  }    \int  {m_a\over p^0  \Big|_{\mathcal P_0}  } 
{ | \vec{p}_\text{\tiny R}| ^2 d| \vec{p}_\text{\tiny R}| d(-\cos\theta_\text{\tiny R})  \Big|_{\mathcal P_0}    \over     \sqrt{  - g_{00} \Big|_{\mathcal P_0}   }  p^0 \Big|_{\mathcal P_0}    }  \\ \nonumber
&\times  {   1 \over   2    | \vec{p}_\text{\tiny R}- \vec{k}_\text{\tiny R}|  \Big|_{\mathcal P_0} }  
\times \delta (  p^0  \Big|_{\mathcal P_0}  - k^0  \Big|_{\mathcal P_0} - { | \vec{p}_\text{\tiny R}- \vec{k}_\text{\tiny R}| \over  \sqrt{ - g_{00} } } \Big|_{\mathcal P_0} )  \\ \nonumber
&\times \{   f_a  ( | \vec{p}_\text{\tiny R} |  \Big|_{\mathcal P_0} )   [ 1+ f_\lambda ( | \vec{k}_\text{\tiny R} |  \Big|_{\mathcal P_0}  )  
   + f_\lambda ( | \vec{p}_\text{\tiny R}- \vec{k}_\text{\tiny R}| \Big|_{\mathcal P_0}  )  ]  \\ \nonumber
& - f_\lambda ( | \vec{k}_\text{\tiny R} |   \Big|_{\mathcal P_0} )   f_\lambda ( | \vec{p}_\text{\tiny R}- \vec{k}_\text{\tiny R}| \Big|_{\mathcal P_0} )  \} ~.
\end{flalign}

Equation \eqref{mass-e-p} at point $\mathcal P_0$ reads 
\begin{flalign} \nonumber
   m^2 +   {\mathcal G}_{ij} p_\text{\tiny R}^i p_\text{\tiny R}^j    \Big|_{\mathcal P_0} 
=m^2 +  | \vec{p}_\text{\tiny R} |^2  \Big|_{\mathcal P_0}  =  - g_{00} (p^0)^2 \Big|_{\mathcal P_0} ~,
\end{flalign}
which gives a differential relation at $\mathcal P_0$
\begin{flalign} \nonumber
| \vec{p}_\text{\tiny R} |  d | \vec{p}_\text{\tiny R} | \Big|_{\mathcal P_0} 
=  - g_{00}   p^0    d p^0  \Big|_{\mathcal P_0}  ~.
\end{flalign}

From the law of cosines \eqref{law-cos}, there is another differential relation which tells us how the magnitude $ | \vec{p}_\text{\tiny R}- \vec{k}_\text{\tiny R}|  $ changes when we change the angle $\theta_\text{\tiny R}$ formed by $\vec{p}_\text{\tiny R}$ and $\vec{k}_\text{\tiny R}$ at $\mathcal P_0$,
\begin{flalign}  \nonumber
 | \vec{p}_\text{\tiny R}- \vec{k}_\text{\tiny R}|   d  | \vec{p}_\text{\tiny R}- \vec{k}_\text{\tiny R}|  \Big|_{\mathcal P_0}  =  | \vec{p}_\text{\tiny R} |     | \vec{k}_\text{\tiny R} |   d ( - \cos\theta_\text{\tiny R} )  \Big|_{\mathcal P_0}  ~.
\end{flalign}

Replacing $  d ( - \cos\theta_\text{\tiny R} ) $ with $d  | \vec{p}_\text{\tiny R}- \vec{k}_\text{\tiny R}|  $, $ | \vec{p}_\text{\tiny R} |  d | \vec{p}_\text{\tiny R} |  $ with  $ - g_{00}   p^0    d p^0  $, and changing the argument of the $\delta$ function, we have
\begin{flalign} \nonumber
&  2 k^0 \Big|_{\mathcal P_0}  \frac{d f_\lambda ( | \vec{k}_\text{\tiny R} |  \Big|_{\mathcal P_0} , \mathcal P_0 )  }{dt} \\ \nonumber
=& \frac{4 m_a \Gamma_{a} }{   ( - g_{00} \Big|_{\mathcal P_0} )  }    \int  {m_a\over p^0  \Big|_{\mathcal P_0}  } 
{  ( - g_{00})   p^0    d p^0  \quad  | \vec{p}_\text{\tiny R}- \vec{k}_\text{\tiny R}|   d  | \vec{p}_\text{\tiny R}- \vec{k}_\text{\tiny R}|     \Big|_{\mathcal P_0}    \over     \sqrt{  - g_{00} \Big|_{\mathcal P_0}   }  p^0 \Big|_{\mathcal P_0}  
 | \vec{k}_\text{\tiny R} |  \Big|_{\mathcal P_0}   }  \\ \nonumber
&\times  {   1 \over   2    | \vec{p}_\text{\tiny R}- \vec{k}_\text{\tiny R}|  \Big|_{\mathcal P_0} }  
\times \sqrt{ - g_{00} }  \Big|_{\mathcal P_0}  \\ \nonumber
&\times    \delta ( | \vec{p}_\text{\tiny R}- \vec{k}_\text{\tiny R}|   \Big|_{\mathcal P_0} - \sqrt{ - g_{00} }  ( p^0   - k^0 ) \Big|_{\mathcal P_0}  )  \\ \nonumber
&\times \{   f_a  ( | \vec{p}_\text{\tiny R} |  \Big|_{\mathcal P_0} )   [ 1+ f_\lambda ( | \vec{k}_\text{\tiny R} |  \Big|_{\mathcal P_0}  )  
   + f_\lambda ( | \vec{p}_\text{\tiny R}- \vec{k}_\text{\tiny R}| \Big|_{\mathcal P_0}  )  ]  \\ \nonumber
& - f_\lambda ( | \vec{k}_\text{\tiny R} |   \Big|_{\mathcal P_0} )   f_\lambda ( | \vec{p}_\text{\tiny R}- \vec{k}_\text{\tiny R}| \Big|_{\mathcal P_0} )  \} ~.
\end{flalign}
After canceling common factors in numerators and denominators, integrating the $| \vec{p}_\text{\tiny R}- \vec{k}_\text{\tiny R}|$ part, the equation becomes
\begin{flalign} \nonumber
&  k^0 \Big|_{\mathcal P_0}  \frac{d f_\lambda ( | \vec{k}_\text{\tiny R} |  \Big|_{\mathcal P_0} , \mathcal P_0 )  }{dt} 
=  m_a \Gamma_{a}    \int  {m_a\over p^0  \Big|_{\mathcal P_0}  } 
{      d p^0    \Big|_{\mathcal P_0}    \over    
 | \vec{k}_\text{\tiny R} |  \Big|_{\mathcal P_0}   }  \\ \nonumber
\times & \{   f_a  ( | \vec{p}_\text{\tiny R} |  \Big|_{\mathcal P_0} )   [ 1+ f_\lambda ( | \vec{k}_\text{\tiny R} |  \Big|_{\mathcal P_0}  )  
   + f_\lambda \left(\sqrt{ - g_{00} }  ( p^0   - k^0 ) \Big|_{\mathcal P_0} \right)  ]  \\ \nonumber
 - & f_\lambda ( | \vec{k}_\text{\tiny R} |   \Big|_{\mathcal P_0} )   f_\lambda \left( \sqrt{ - g_{00} }  ( p^0   - k^0 ) \Big|_{\mathcal P_0} \right)  \} ~.
\end{flalign}
The rate of change of photon occupation number at ${\mathcal P_0}$ is then
\begin{flalign} \nonumber
&    \frac{d f_\lambda ( | \vec{k}_\text{\tiny R} |  \Big|_{\mathcal P_0} , \mathcal P_0 )  }{dt} 
=  { m_a \Gamma_{a} \over  k^0 | \vec{k}_\text{\tiny R} |  \Big|_{\mathcal P_0}  }   \int  {m_a\over p^0  \Big|_{\mathcal P_0}  } 
d p^0    \Big|_{\mathcal P_0}   \\ \nonumber
\times & \{   f_a  ( | \vec{p}_\text{\tiny R} |  \Big|_{\mathcal P_0} )   [ 1+ f_\lambda ( | \vec{k}_\text{\tiny R} |  \Big|_{\mathcal P_0}  )  
   + f_\lambda \left(\sqrt{ - g_{00} }  ( p^0   - k^0 ) \Big|_{\mathcal P_0} \right)  ]  \\ \nonumber
 - & f_\lambda ( | \vec{k}_\text{\tiny R} |   \Big|_{\mathcal P_0} )   f_\lambda \left( \sqrt{ - g_{00} }  ( p^0   - k^0 ) \Big|_{\mathcal P_0} \right)  \} ~.
\end{flalign}

Now the rate equation is given at event $\mathcal P_0(t_{\mathcal P_0}, x^i_{\mathcal P_0})$ where the metric is Euclidean, using Riemann normal coordinates $x_\text{\tiny R}^i$ on the 3-surface. The norms of 3-momenta keep the same value, regardless of which coordinates were used. We can converte from Riemann normal coordinates $x_\text{\tiny R}^i$ back to general coordinates, via the following substitutions
\begin{flalign}  \nonumber
\sqrt{ g_{ij} p^i p^j }  \Big|_{\mathcal P_0}
&\quad\Leftarrow   \sqrt{ {\mathcal G}_{ij} p_\text{\tiny R}^i p_\text{\tiny R}^j }  \Big|_{\mathcal P_0} 
=  | \vec{p}_\text{\tiny R} |  \Big|_{\mathcal P_0}  \\\nonumber
\sqrt{ g_{ij} k^i k^j }  \Big|_{\mathcal P_0}
&\quad\Leftarrow   \sqrt{ {\mathcal G}_{ij} k_\text{\tiny R}^i k_\text{\tiny R}^j }  \Big|_{\mathcal P_0} 
=  | \vec{k}_\text{\tiny R} |  \Big|_{\mathcal P_0}  
\end{flalign}
\begin{flalign}  \nonumber
  \sqrt{ g_{ij}  (p^i - k^i) (p^j - k^j)} \Big|_{\mathcal P_0}   
\quad\Leftarrow  & \sqrt{ {\mathcal G}_{ij}  (p_\text{\tiny R}^i-k_\text{\tiny R}^i) (p_\text{\tiny R}^j-k_\text{\tiny R}^j)} \Big|_{\mathcal P_0} 
=  | \vec{p}_\text{\tiny R}- \vec{k}_\text{\tiny R}|  \Big|_{\mathcal P_0}
\end{flalign}
and
\begin{flalign}  \nonumber
& k^0  \Big|_{\mathcal P_0}  = \sqrt{ g_{ij} k^i k^j \over  - g_{00}  }   \Big|_{\mathcal P_0}  ~.
\end{flalign}

The reason we can use Riemann normal coordinate and convert back is that, the equations do not depend on the specific  Riemann normal coordinates we are using, and where the event point is located. All the relevant variables can be written in covariant form. We can go through the same process at event $\mathcal P_1(t_{\mathcal P_1}, x^i_{\mathcal P_1})$, $\mathcal P_2(t_{\mathcal P_2}, x^i_{\mathcal P_2})$ using Riemann normal coordinates $x_\text{\tiny R1}^i$ and $x_\text{\tiny R2}^i$ respectively, and then obtain equations of the same form. Thus we arrive at
\begin{flalign} \nonumber
&    \frac{d f_\lambda ( \sqrt{ g_{ij} k^i k^j }   \Big|_{\mathcal P_0} , \mathcal P_0 )  }{dt} 
=  { m_a \Gamma_{a} \sqrt{   - g_{00}  }\Big|_{\mathcal P_0}  \over    g_{ij} k^i k^j \Big|_{\mathcal P_0}  }   \int  {m_a\over p^0  \Big|_{\mathcal P_0}  } 
d p^0    \Big|_{\mathcal P_0}   \\ \nonumber
& \times  \{   f_a  ( \sqrt{ g_{ij} p^i p^j }   \Big|_{\mathcal P_0} )   [ 1+ f_\lambda ( \sqrt{ g_{ij} k^i k^j }   \Big|_{\mathcal P_0}  )   \\ \nonumber
&   + f_\lambda \left(\sqrt{ - g_{00} }  ( p^0   - k^0 ) \Big|_{\mathcal P_0} \right)  ]  \\ \nonumber
 &  -  f_\lambda (  \sqrt{ g_{ij} k^i k^j }    \Big|_{\mathcal P_0} )   f_\lambda \left( \sqrt{ - g_{00} }  ( p^0   - k^0 ) \Big|_{\mathcal P_0} \right)  \} ~.
\end{flalign}
If the event $\mathcal P_0(t_{\mathcal P_0}, x^i_{\mathcal P_0})$ is not at a special point in spacetime, then we should have this equation \eqref{rate-equa-5} at any location $x^\alpha$,
\begin{flalign} \nonumber
&    \frac{d f_\lambda ( \sqrt{ g_{ij} k^i k^j }   )  }{dt}   \\ \nonumber
= &  { m_a \Gamma_{a} \sqrt{   - g_{00}  }  \over    g_{ij} k^i k^j   }   \int  {m_a\over p^0    } 
d p^0   \times  \{   f_a  ( \sqrt{ g_{ij} p^i p^j }    )   \\ \nonumber
&   [ 1+ f_\lambda ( \sqrt{ g_{ij} k^i k^j }     )   + f_\lambda \left(\sqrt{ - g_{00} }  ( p^0   - k^0 )   \right)  ]  \\ \nonumber
 &  -  f_\lambda (  \sqrt{ g_{ij} k^i k^j }      )   f_\lambda \left( \sqrt{ - g_{00} }  ( p^0   - k^0 )   \right)  \} ~.
\end{flalign}
We note the factor of $\sqrt{   - g_{00}  }$ is from the gravitational redshift which corrects the time difference between the clock in the lab and the clock at the location the axion. The factor $ {m_a\over p^0}$ is due to special relativity correction.\\

\subsection{Maximum momentum of axion}

An axion at $r$ has momentum $p^\alpha$,
\begin{flalign}  \nonumber
m_a^2 + g_{ij}(r) p^i p^j  =&  - g_{00}(r) (p^0)^2  ~.
\end{flalign}
In our setup of simple model, we placed axions and photons inside the region $\Theta(r_+ - r)\Theta(r-r_-)$.  If the axion moves on a geodesic, then the scalar product of its momentum and the killing vector $(1,0,0,0)$ is a constant,
\begin{flalign}  \nonumber
g_{00}(r) p^0(r)=g_{00}( r_+) p^0( r_+)  ~.
\end{flalign}
The momentum is  normalized at $ r_+$ such that
\begin{flalign}  \nonumber
m_a^2 + g_{ij}( r_+) p^i( r_+)   p^j ( r_+)   =&  - g_{00}( r_+) [ p^0 ( r_+) ]^2   ~.
\end{flalign}
 Considering an axion with just enough momentum to reach radius $ r_+$ where $p^i ( r_+)=0$, then we have
\begin{flalign}  \nonumber
p^0_\text{esc}( r_+) = { m_a \over \sqrt{ - g_{00}( r_+) }  } ~.
\end{flalign}
Hence the escape value of  $p^0$ for an axion at any $r < r_+$ is
\begin{flalign}  \nonumber
p^0_\text{esc}(r)= { g_{00}( r_+) \over  g_{00} (r) }  p^0_\text{esc}( r_+) = { \sqrt{ - g_{00}( r_+) } \over - g_{00}(r)   } m_a   ~.
\end{flalign}
The escape momentum $\sqrt{g_{ij}(r) p^i p^j}_\text{esc}$ thus satisfy
\begin{flalign}  \nonumber
m_a^2 + [ g_{ij}(r) p^i p^j ]_\text{esc}   =&   { - g_{00}( r_+) \over - g_{00}(r) } m_a^2   ~,
\end{flalign}
or upon  solving for this momentum,
\begin{flalign}  \nonumber
\sqrt{g_{ij}(r) p^i p^j}_\text{esc}  =&  m_a  \sqrt{ {- g_{00}( r_+) \over - g_{00} (r)} - 1 }   ~.
\end{flalign}
The escape momenta are different for axions at different $r$, and the maximally allowed momentum should be set to the largest $\sqrt{g_{ij}(r) p^i p^j}_\text{esc}$.
\begin{flalign}  \nonumber
p_\text{max}  =m_a \beta' =   \text{max}\{  m_a  \sqrt{ {- g_{00}( r_+)\over - g_{00} (r)} - 1 }  \}  ~,
\end{flalign}
where we have defined a relativistic parameter $\beta'$. In our setup of the simple model, 
\begin{flalign}  \label{beta-static}
 \beta' =    \sqrt{ {- g_{00}( r_+)\over - g_{00} (r_-)} - 1 }   ~.
\end{flalign}

For Schwarzschild spacetime, the  relativistic parameter or  the maximum velocity is
\begin{flalign}  \nonumber
\beta'_\text{sch}  =&   \sqrt{  {2M\over r_- } -{2M\over r_+ } \over 1-{2M\over r_- } }    \gtrsim \sqrt{ {2M\over r_- } -{2M\over r_+ } } = \sqrt{ 2 M \over r_+} \sqrt{ {r_+ \over r_- } -1 } ~.
\end{flalign}
Using solar parameters, 
the numerical value of the maximum velocity is written as
\begin{flalign}  \nonumber
\beta'_\text{sch}  =&  2\times 10^{-3} \sqrt{\frac{M R_\odot}{ M_\odot r_+ }}
\sqrt{ {r_+ \over r_- } -1 } ~.
\end{flalign}
Depending on the outer radius of the cluster and the mass of the Schwarzschild host, there may be a noticable correction to the value of maximum velocity obtained from Newtonian theory. The maximum velocity $\beta$ of non-relativistic axion calculated in \cite{Kephart:1994uy} is from self gravity, here $\beta'$ is the result of gravitational bound from the host. \\

Equation \eqref{beta-static} is a general formula for calculating the upper bound on $\sqrt{ g_{ij} p^i p^j }$ without specifically knowing each individual component of the metric $g_{ij}$ of the static spacetime.
By itself equation \eqref{beta-static} can not always constrain axions to be inside $\Theta(r_+ - r)\Theta(r-r_-)$. 
Axions starting with $\sqrt{ g_{ij} p^i p^j }=m_a\beta'$ at $r=r_-$ and at $r=r_+$ can barely reach $r_+$ and $r_\text{max}\approx r_+$, respectively. 
The point is we can always find the maximum momentum of axions in the region, but the exact formula for the relativistic parameter $\beta'$ is not that important.
With any nonzero angular momentum, $r_-\gg 2M$ limits the possibility of incidents that particles falling into the host.\\

\subsection{ Integration over $|k_1^i|$ $|k^i|$}
We choose not to impose the approximations on eq. \eqref{rate-equa-2}   used in \cite{Kephart:1994uy}. There the axions were treated as non-relativistic, so the maximum momentum of the axion $m_a\gamma\beta$ became $m_a\beta$. If we imposed this condition here, then due to the small value of $\beta'$ derived previously, we find $\sqrt{ - g_{00} }  p^0= \sqrt{ m_a^2 + g_{ij}  p^i p^j  } \sim m_a$, and \eqref{rate-equa-2} would become,
\begin{flalign} \nonumber
&   \frac{d n_\lambda  }{dt}  =    { m_a \Gamma_{a} \sqrt{   - g_{00}  }  \over   2\pi^2   }  \int  \int_{  m_a^2 \over 4 \sqrt{ g_{ij} k^i k^j }   }   d ( \sqrt{ g_{ij} k_1^i k_1^j })   d( \sqrt{ g_{ij} k^i k^j } )    \\ \nonumber
& \times  \{   f_a  \left( \sqrt{   ( \sqrt{ g_{ij} k^i k^j } + \sqrt{ g_{ij} k_1^i k_1^j })^2 - m_a^2 }    \right)  [ 1+ f_\lambda ( \sqrt{ g_{ij} k^i k^j }     )    \\ \nonumber
&    + f_\lambda ( \sqrt{ g_{ij} k_1^i k_1^j }   )  ]     -  f_\lambda (  \sqrt{ g_{ij} k^i k^j }      )   f_\lambda ( \sqrt{ g_{ij} k_1^i k_1^j }   )  \}     ~.
\end{flalign}
This approximation drops the special relativity correction, which account for the time difference between the clock of a stationary observer at $x^i$ and the intrinsic clock of the axion. 

But since we have decided to keep this time dilation factor, we start with \eqref{rate-equa-2}.
\begin{flalign} \nonumber
&   \frac{d n_\lambda  }{dt}    \\ \nonumber
= &  { m_a \Gamma_{a} \sqrt{   - g_{00}  }  \over   2\pi^2   }  \int  \int_{  m_a^2 \over 4 \sqrt{ g_{ij} k^i k^j }   }    {m_a\over  \sqrt{ g_{ij} k^i k^j } +  \sqrt{ g_{ij} k_1^i k_1^j }   }  \\ \nonumber
& \times  \{   f_a  \left( \sqrt{   ( \sqrt{ g_{ij} k^i k^j } + \sqrt{ g_{ij} k_1^i k_1^j })^2 - m_a^2 }    \right)  [ 1+ f_\lambda ( \sqrt{ g_{ij} k^i k^j }     )    \\ \nonumber
&    + f_\lambda ( \sqrt{ g_{ij} k_1^i k_1^j }   )  ]     -  f_\lambda (  \sqrt{ g_{ij} k^i k^j }      )   f_\lambda ( \sqrt{ g_{ij} k_1^i k_1^j }   )  \}   \\ \nonumber
& \times  d ( \sqrt{ g_{ij} k_1^i k_1^j })   d( \sqrt{ g_{ij} k^i k^j } )   ~.
\end{flalign}

The first integral is
\begin{flalign} \nonumber
& \int d( \sqrt{ g_{ij} k^i k^j } ) \int_{  m_a^2 \over 4 \sqrt{ g_{ij} k^i k^j } }  d( \sqrt{ g_{ij} k_1^i k_1^j })  \\ \nonumber
&  {m_a\over  \sqrt{ g_{ij} k^i k^j } +  \sqrt{ g_{ij} k_1^i k_1^j }   }
 f_a  \left( \sqrt{   ( \sqrt{ g_{ij} k^i k^j } + \sqrt{ g_{ij} k_1^i k_1^j })^2 - m_a^2 }    \right) \\ \nonumber
=& \int d( \sqrt{ g_{ij} k^i k^j } )           \int^{m_a\sqrt{1+\beta'^2}  - \sqrt{ g_{ij} k^i k^j }}_{m_a^2  \over 4  \sqrt{ g_{ij} k^i k^j }  }    d( \sqrt{ g_{ij} k_1^i k_1^j })  \\ \nonumber
&   {m_a\over  \sqrt{ g_{ij} k^i k^j } +  \sqrt{ g_{ij} k_1^i k_1^j }   }   [ f_{ac}(t)\Theta(r_+ - r)\Theta(r-r_-)   + f_{ad}(t) d(r) ]   \\ \nonumber
=& \int d( \sqrt{ g_{ij} k^i k^j } )  [ f_{ac}(t)\Theta(r_+ - r)\Theta(r-r_-)   + f_{ad}(t) d(r) ]   \\ \nonumber
&\times  m_a \ln( \sqrt{ g_{ij} k^i k^j } +  \sqrt{ g_{ij} k_1^i k_1^j }  )  \bigg|^{m_a\sqrt{1+\beta'^2}  - \sqrt{ g_{ij} k^i k^j }}_{m_a^2  \over 4  \sqrt{ g_{ij} k^i k^j }  }  \\ \nonumber
=& m_a \int d( \sqrt{ g_{ij} k^i k^j } )   [ f_{ac}(t)\Theta(r_+ - r)\Theta(r-r_-)   + f_{ad}(t) d(r) ]  \\ \nonumber
&\times  [ \ln(   m_a\sqrt{1+\beta'^2}     )   -  \ln( \sqrt{ g_{ij} k^i k^j } +  { m_a^2  \over 4  \sqrt{ g_{ij} k^i k^j } }    )  ]  
  \\ \nonumber\end{flalign}
Now let
\begin{flalign} \nonumber
z =&\sqrt{ g_{ij} k^i k^j }  \\ \nonumber
z_\pm=&\sqrt{ g_{ij} k^i k^j }_\pm ={ m_a \over  2 }  (   \sqrt{1+\beta'^2} \pm  \beta'  )  ~, \\ \nonumber
\end{flalign}
so that the first integral becomes
\begin{flalign} \nonumber
=& m_a \int d z   [ \ln(  m_a\sqrt{1+\beta'^2}   )   -  \ln( z +  { m_a^2  \over 4  z }    )  ] \\ \nonumber
&\times  [ f_{ac}(t)\Theta(r_+ - r)\Theta(r-r_-)   + f_{ad}(t) d(r) ]   \\ \nonumber
=& m_a  [  \int \ln( 4 m_a\sqrt{1+\beta'^2}   z  ) dz  -  \int \ln( 4 z^2 +   m_a^2    ) dz   ] \\ \nonumber
&\times  [ f_{ac}(t)\Theta(r_+ - r)\Theta(r-r_-)   + f_{ad}(t) d(r) ]  \\ \nonumber
=& m_a  \{ [ z \ln( 4 m_a\sqrt{1+\beta'^2}   z  )  - z ] \bigg|^+_-   \\ \nonumber
& -  [ m_a \arctan{2z\over m_a} -2z +z\ln(4 z^2 +   m_a^2)  ]\bigg|^+_-  \} \\ \nonumber
&\times  [ f_{ac}(t)\Theta(r_+ - r)\Theta(r-r_-)   + f_{ad}(t) d(r) ]     \\ \nonumber
=& m_a  \{ [ z \ln( 4 m_a\sqrt{1+\beta'^2}   z  ) - z\ln(4 z^2 +   m_a^2)     \\ \nonumber
& + z  -   m_a \arctan{2z\over m_a}     ]\bigg|^+_-  \} \\ \nonumber
&\times  [ f_{ac}(t)\Theta(r_+ - r)\Theta(r-r_-)   + f_{ad}(t) d(r) ]  
\end{flalign}
where using
\begin{flalign} \nonumber
& [ z \ln( 4 m_a\sqrt{1+\beta'^2}   z  ) ] \bigg|^+_-  \\ \nonumber
=& m_a \beta'   \ln( 2 m_a^2 \sqrt{1+\beta'^2}   )     \\ \nonumber
& +  { m_a \over  2 } \sqrt{1+\beta'^2} \ln[ {    \sqrt{1+\beta'^2} +  \beta'  
	  \over \sqrt{1+\beta'^2} -  \beta'   }  ]  
\end{flalign}
and
\begin{flalign} \nonumber
& z\ln(4 z^2 +   m_a^2) \bigg|^+_-  \\ \nonumber
=& { m_a \over  2 }    \sqrt{1+\beta'^2} \ln[ {  ( \sqrt{1+\beta'^2} +  \beta'  )^2 +   1 \over    ( \sqrt{1+\beta'^2} -  \beta'  )^2 +   1 }]  \\ \nonumber
& + m_a  \beta'   \ln ( 2 m_a^2 \sqrt{  1 +  \beta'^2  }  )
\end{flalign}
as well as
\begin{flalign} \nonumber
 [ z \ln( 4 m_a\sqrt{1+\beta'^2}   z  ) ] - z\ln(4 z^2 +   m_a^2) \bigg|^+_-  =  0
\end{flalign}
and
\begin{flalign} \nonumber
&z  -   m_a \arctan{2z\over m_a}   \bigg|^+_-  \\ \nonumber
=& m_a [ \beta'  +  \arctan(\sqrt{1+\beta'^2} -  \beta') -   \arctan(\sqrt{1+\beta'^2} +  \beta')  ]  \\ \nonumber
=& m_a [ \beta'  + ( {\pi\over4} - {\beta'\over2} + {\beta'^3\over6} - {\beta'^5\over10} +...) \\ \nonumber
   &   -   ( {\pi\over4} + {\beta'\over2} - {\beta'^3\over6} + {\beta'^5\over10} +... )  ]  = m_a    {\beta'^3\over 3}   
\end{flalign}
the first integral finally reduces to
\begin{flalign} \nonumber
& \int d( \sqrt{ g_{ij} k^i k^j } ) \int_{  m_a^2 \over 4 \sqrt{ g_{ij} k^i k^j } }  d( \sqrt{ g_{ij} k_1^i k_1^j })  \\ \nonumber
&  {m_a\over  \sqrt{ g_{ij} k^i k^j } +  \sqrt{ g_{ij} k_1^i k_1^j }   } \\ \nonumber
& f_a  ( \sqrt{   ( \sqrt{ g_{ij} k^i k^j } + \sqrt{ g_{ij} k_1^i k_1^j })^2 - m_a^2 }    ) \\ \nonumber
= &     {  m_a^2 \beta'^3 \over 3}      [ f_{ac}(t)\Theta(r_+ - r)\Theta(r-r_-)   + f_{ad}(t) d(r) ]     ~.
\end{flalign}\\

Using a similar procedure we find that the second integral is
\begin{flalign} \nonumber
& \int d( \sqrt{ g_{ij} k^i k^j } ) \int_{  m_a^2 \over 4 \sqrt{ g_{ij} k^i k^j } }  d( \sqrt{ g_{ij} k_1^i k_1^j })  \\ \nonumber
&  {m_a\over  \sqrt{ g_{ij} k^i k^j } +  \sqrt{ g_{ij} k_1^i k_1^j }   } \\ \nonumber
& f_a  \left( \sqrt{   ( \sqrt{ g_{ij} k^i k^j } + \sqrt{ g_{ij} k_1^i k_1^j })^2 - m_a^2 }    \right)  f_\lambda ( \sqrt{ g_{ij} k^i k^j }     )  \\ \nonumber
= &   \int d( \sqrt{ g_{ij} k^i k^j } )  \int_{ { m_a^2 \over 4 \sqrt{ g_{ij} k^i k^j } }}^{m_a\sqrt{1+\beta'^2}  - \sqrt{ g_{ij} k^i k^j } }  d( \sqrt{ g_{ij} k_1^i k_1^j })       \\ \nonumber
&\times   {m_a\over  \sqrt{ g_{ij} k^i k^j } +  \sqrt{ g_{ij} k_1^i k_1^j }   } \\ \nonumber
&\times    \{  f_{ac}(t)  f_{\lambda c}(t)\Theta(r_+ - r)\Theta(r-r_-)     \\ \nonumber
& + [ ( f_{ac}(t)  f_{\lambda d}(t)  +  f_{ad}(t)  f_{\lambda c}(t)  )  d(r) ]  \}  \\ \nonumber
= & m_a  \{ [ z \ln( 4 m_a\sqrt{1+\beta'^2}   z  ) - z\ln(4 z^2 +   m_a^2)     \\ \nonumber
& + z  -   m_a \arctan{2z\over m_a}     ]\bigg|^+_-  \} \\ \nonumber
&\times     \{  f_{ac}(t)  f_{\lambda c}(t)\Theta(r_+ - r)\Theta(r-r_-)     \\ \nonumber
& + [  f_{ac}(t)  f_{\lambda d}(t)  +  f_{ad}(t)  f_{\lambda c}(t)  ]  d(r) \}  \\ \nonumber
= & {  m_a^2 \beta'^3 \over 3}      \{  f_{ac}(t)  f_{\lambda c}(t)\Theta(r_+ - r)\Theta(r-r_-)     \\ \nonumber
& + [  f_{ac}(t)  f_{\lambda d}(t)  +  f_{ad}(t)  f_{\lambda c}(t)  ]  d(r) \}  ~.
\end{flalign}
Here terms which are second or higher order of the distortion $d(r)$ are dropped. Because the distortion is expected to be small when the locations of all the interactions are far away from the event horizon.

Likewise, the third integral is
\begin{flalign} \nonumber
& \int d( \sqrt{ g_{ij} k^i k^j } ) \int_{  m_a^2 \over 4 \sqrt{ g_{ij} k^i k^j } }  d( \sqrt{ g_{ij} k_1^i k_1^j })  \\ \nonumber
& {m_a\over  \sqrt{ g_{ij} k^i k^j } +  \sqrt{ g_{ij} k_1^i k_1^j }   } \\ \nonumber
& f_a  \left( \sqrt{   ( \sqrt{ g_{ij} k^i k^j } + \sqrt{ g_{ij} k_1^i k_1^j })^2 - m_a^2 }    \right)  f_\lambda ( \sqrt{ g_{ij} k_1^i k_1^j }      )  \\ \nonumber
= &   m_a  \{ [ z \ln( 4 m_a\sqrt{1+\beta'^2}   z  ) - z\ln(4 z^2 +   m_a^2)     \\ \nonumber
& + z  -   m_a \arctan{2z\over m_a}     ]\bigg|^+_-  \}  \\ \nonumber
&\times     \{  f_{ac}(t)  f_{\lambda c}(t)\Theta(r_+ - r)\Theta(r-r_-)     \\ \nonumber
& + [  f_{ac}(t)  f_{\lambda d}(t)  +  f_{ad}(t)  f_{\lambda c}(t)  ]  d(r) \}  \\ \nonumber
= &   {  m_a^2 \beta'^3 \over 3}    \{  f_{ac}(t)  f_{\lambda c}(t)\Theta(r_+ - r)\Theta(r-r_-)     \\ \nonumber
& + [  f_{ac}(t)  f_{\lambda d}(t)  +  f_{ad}(t)  f_{\lambda c}(t)  ]  d(r) \}  ~.
\end{flalign}
The last integral splits into two parts: back reactions produce normal axions with $\sqrt{g_{ij}(r) p^i p^j} \leq m_a\beta'$ and sterile axions with $\sqrt{g_{ij}(r) p^i p^j} > m_a\beta'$.
\begin{flalign} \nonumber
& \int d( \sqrt{ g_{ij} k^i k^j } ) \int_{  m_a^2 \over 4 \sqrt{ g_{ij} k^i k^j } }  d( \sqrt{ g_{ij} k_1^i k_1^j })  \\ \nonumber
& \times  {m_a\over  \sqrt{ g_{ij} k^i k^j } +  \sqrt{ g_{ij} k_1^i k_1^j }   }  
  f_\lambda (  \sqrt{ g_{ij} k^i k^j }      )   f_\lambda ( \sqrt{ g_{ij} k_1^i k_1^j }   )  \\ \nonumber
=&  \int d( \sqrt{ g_{ij} k^i k^j } ) \int_{  m_a^2 \over 4 \sqrt{ g_{ij} k^i k^j } }^{m_a\sqrt{1+\beta'^2}  - \sqrt{ g_{ij} k^i k^j } }  d( \sqrt{ g_{ij} k_1^i k_1^j })   \\ \nonumber
& \times  {m_a\over  \sqrt{ g_{ij} k^i k^j } +  \sqrt{ g_{ij} k_1^i k_1^j }   }   +  {m_a\over  \sqrt{ g_{ij} k^i k^j } +  \sqrt{ g_{ij} k_1^i k_1^j }   }   \\ \nonumber
 & \times \int d( \sqrt{ g_{ij} k^i k^j } ) \int_{m_a\sqrt{1+\beta'^2}  - \sqrt{ g_{ij} k^i k^j } }^{\sqrt{ g_{ij} k^i k^j }_+}   d( \sqrt{ g_{ij} k_1^i k_1^j })     \\ \nonumber
& \times  [ f_{\lambda c}^2(t)\Theta(r_+ - r)\Theta(r-r_-) + 2 f_{\lambda c}(t) f_{\lambda d}(t) d(r)   ]  \\ \nonumber
=&  m_a^2 (    {\beta'^3\over 3} +  {\beta'^2 \over2}  ) [ f_{\lambda c}^2(t)\Theta(r_+ - r)\Theta(r-r_-)   \\ \nonumber
&+ 2 f_{\lambda c}(t) f_{\lambda d}(t) d(r)   ] 
\end{flalign}
The integral which accounts for the sterile axions is worked out as follows,
\begin{flalign} \nonumber
& \int d( \sqrt{ g_{ij} k^i k^j } ) \int_{m_a\sqrt{1+\beta'^2}  - \sqrt{ g_{ij} k^i k^j } }^{\sqrt{ g_{ij} k^i k^j }_+}   d( \sqrt{ g_{ij} k_1^i k_1^j })     \\ \nonumber
& \times  {m_a\over  \sqrt{ g_{ij} k^i k^j } +  \sqrt{ g_{ij} k_1^i k_1^j }   }     \\ \nonumber
=&  m_a  \int d( \sqrt{ g_{ij} k^i k^j } )     [    \ln( \sqrt{ g_{ij} k^i k^j } +  { m_a \over  2 }  (   \sqrt{1+\beta'^2} +  \beta'  )  )   
\\ \nonumber
 & - \ln(  m_a\sqrt{1+\beta'^2}    )      ] \\ \nonumber
=& m_a  \{  [ z+ { m_a \over  2 }  (   \sqrt{1+\beta'^2} +  \beta'  ) ]  \ln[ z +  { m_a \over  2 }  (   \sqrt{1+\beta'^2} +  \beta'  ) ] \\ \nonumber
&     -z  - z \ln(  m_a\sqrt{1+\beta'^2}    )        \}\bigg|^+_-   \\ \nonumber
=& m_a  \{  m_a (   \sqrt{1+\beta'^2} +  \beta'  )   \ln [ m_a   (   \sqrt{1+\beta'^2} +  \beta'  )     ] \\ \nonumber
& - m_a   \sqrt{1+\beta'^2}    \ln [ m_a     \sqrt{1+\beta'^2}  ]  \\ \nonumber
& -  m_a\beta' - m_a\beta'  \ln(  m_a\sqrt{1+\beta'^2}    )  \}\\ \nonumber
=& m_a^2  [ (   \sqrt{1+\beta'^2} +  \beta'  )   \ln    (  {  \sqrt{1+\beta'^2} +  \beta' \over  \sqrt{1+\beta'^2} } ) - \beta' ]  \\ \nonumber
=& m_a^2 [ ( \beta' + {\beta'^2 \over2} - {\beta'^3 \over6}+... )  - \beta'  ] \approx m_a^2 {\beta'^2 \over2} ~.
\end{flalign}

Substituting all  four integrals into rate equation, we finally arrive at \eqref{DnPhotonDt}
\begin{flalign} \nonumber
&  \frac{d n_\lambda   }{dt} =  { m_a \Gamma_{a} \over   2\pi^2   }  \sqrt{   - g_{00}  }    \\ \nonumber
&\times   \{  {  m_a^2 \beta'^3 \over 3}  [  f_{ac}(t)\Theta(r_+ - r)\Theta(r-r_-)   + f_{ad}(t) d(r)    \\ \nonumber
 &     +  2 f_{ac}(t)  f_{\lambda c}(t)\Theta(r_+ - r)\Theta(r-r_-)    \\ \nonumber
 & + 2  (  f_{ac}(t)  f_{\lambda d}(t)  +  f_{ad}(t)  f_{\lambda c}(t)  ) d(r)   \\ \nonumber
& -   f_{\lambda c}^2(t)\Theta(r_+ - r)\Theta(r-r_-) - 2 f_{\lambda c}(t) f_{\lambda d}(t) d(r)   ]  \\ \nonumber
&  -    { m_a^2\beta'^2 \over  2 }       [ f_{\lambda c}^2(t)\Theta(r_+ - r)\Theta(r-r_-) + 2 f_{\lambda c}(t) f_{\lambda d}(t) d(r)   ]   \}   ~.
\end{flalign}

\vfill

\begin{acknowledgments} We thank Kevin Ludwick for a helpful correspondence.
This work  was supported by US DOE grant DE-SC0019235.
 
\end{acknowledgments}

\end{document}